\documentclass{article}

\usepackage[utf8]{inputenc} % allow utf-8 input
\usepackage[T1]{fontenc}    % use 8-bit T1 fonts
\usepackage{hyperref}       % hyperlinks
\usepackage{url}            % simple URL typesetting
\usepackage{booktabs}       % professional-quality tables
\usepackage{amsfonts}       % blackboard math symbols
\usepackage{nicefrac}       % compact symbols for 1/2, etc.
\usepackage{microtype}      % microtypography
\usepackage{comment}

\usepackage{amsmath,amsfonts,amsthm,amssymb}

\newtheorem{propo}{Proposition}[section]
\newtheorem{lemma}[propo]{Lemma}

\newtheorem{coro}[propo]{Corollary}
\newtheorem{thm}{Theorem}

\newtheorem{ansatz}{Ansatz}

\usepackage{graphicx,float,wrapfig}
\usepackage{multirow}

\def\hf{\widehat{f}}

\def\reals{{\mathbb R}}

\def\<{\langle}
\def\>{\rangle}
\def\E{{\mathbb E}}

\def\hH{{\widehat{H}}}

\def\hJ{{\widehat{J}}}
\def\cS{{\cal T}}
\def\bh{h}
\def\tS{{\tilde{S}}}
\def\tA{{\widetilde{A}}}
\def\Pr{{\mathbb P}}

\def\Var{{\textrm{ Var }}}

\title{
Density Functional Estimators with 
$k$-Nearest Neighbor Bandwidths
}

\author{
 Weihao Gao\thanks{Coordinated Science Lab and Department of Electrical and Computer Engineering}, \;\;
 Sewoong Oh\thanks{Coordinated Science Lab and Department of Industrial and Enterprise Systems Engineering}, \;\;
 and \;\;Pramod Viswanath$^*$\\
   University of Illinois at Urbana-Champaign\\
  Urbana, IL 61801  \\
  \texttt{\{wgao9,swoh,pramodv\}@illinois.edu} \\
}
\date{}

\begin{document}

\maketitle

\begin{abstract}
Estimating  expected polynomials of density functions from samples is 
a basic problem with numerous applications in statistics and information theory. 
Although kernel density estimators are widely used in practice for such functional estimation problems, practitioners 
are left on their own to choose an appropriate bandwidth for each application in hand. 
Further, kernel density estimators suffer from boundary biases, which are prevalent in real world data with lower dimensional structures. 
We propose  using the fixed-$k$ nearest neighbor distances for the bandwidth, which  adaptively adjusts to local geometry. 
Further, we propose  a novel estimator based on {\em local likelihood density estimators}, 
 that mitigates  the boundary biases.  
 Although such a choice of fixed-$k$ nearest neighbor distances to bandwidths  results in 
 inconsistent estimators, 
 we provide a simple debiasing scheme that precomputes the asymptotic bias and divides off this term. 
With this novel correction, we show consistency of this debiased estimator. 
We  provide numerical experiments suggesting that it improves upon competing state-of-the-art methods. 
\end{abstract}

\section{Introduction}

Estimating unknown distributions (probability mass functions (pmf) for discrete alphabets or probability density functions (pdf) for continuous alphabets) based on observed samples is one of the most important problem in statistics.
%This has been studied for decades and has numerous successful downstream applications.
In this paper, we address the problem of estimating expectations of {\em functionals} of the density  from samples.
%, rather than the pmf or pdf alone.
For discrete random variables, recent works \cite{HJW15,WY16,BZLV16}  use polynomial approximations
of the  functionals to trade-off the bias and variance and achieve
the minimax optimal rate for the problem of estimating entropic quantities, such as the Shannon entropy, the mutual information
and the Kullback-Leibler divergence.

Motivated by recent advances in the discrete case, in this paper we investigate the continuous setting, 
focusing on  the problem of estimating the integral of a {\em polynomial functional} of density function. We are particularly interested in  high-dimensional settings, where the problem is both of practical interest as well as technically challenging.
Polynomial functionals are linear combinations of {\em monomial functionals}, and it suffices to study the estimation of integral of monomial functionals of density, i.e., 
\begin{eqnarray}
\label{eq:integ}
J_{\alpha}(X) &\equiv& \int_{\mathbb{R}^d} f^{\alpha}(x) dx \;,
\end{eqnarray}
for $\alpha \neq 1$ (for $\alpha =1 $, the integral is simply $1$). Note that such an estimator  immediately provides an estimate of   the R\'{e}nyi entropy~\cite{Renyi61} defined as $H_{\alpha}(X) \equiv \log \int_{\mathbb{R}^d} f^{\alpha}(x) dx/(1-\alpha) = \log J_{\alpha}(X)/(1-\alpha)$. The R\'{e}nyi entropy can be estimated as $\hH_{\alpha}(X) = \log \hJ_{\alpha}(X)/(1-\alpha)$, which has immediate  applications in several problems such as fractal random walks~\cite{AZ94}, image registration and indexing, texture classification and image matching~\cite{Hero99,Hero02,Hero05}, and parameter estimation in semi-parametric models~\cite{WTP05}.

The most widely used estimators of $J_{\alpha}(X)$ are the so called {\em resubstitution estimators}. These estimators are based on the fact that $J_{\alpha}(X) = \E \left[\, f^{\alpha-1}(X) \,\right]$. Given any density estimator $\widehat{f}$, one can substitute the integration by a sample mean and use $\hJ^{(n)}_{\alpha}(X) = (1/n)\sum_{i=1}^n \widehat{f}^{\alpha-1}(X_i)$. While any density estimator can be used to develop a resubstitution estimator, the state-of-the-art estimators~\cite{Joe89,Leo08,Pal10} use either kernel density estimators (KDE) or $k$-nearest-neighbor ($k$-NN) methods. According to~\cite{Joe89}, the KDE based estimator achieves the minimax optimal convergence rate given in~\cite{BM95} for $d \leq 6$, for a class of smooth-enough densities. The $k$-NN based estimator achieves the minimax rate for $d \leq 2$ according to~\cite{GOV16De} (or for $d\leq 4$ under certain smoothness assumptions on the distribution in \cite{BSY16}).

Despite the theoretical guarantees of the aforementioned estimators, they still suffer in practical applications, especially when the dimension might be  large.
%The problem is that these methods assume that the density is smooth enough around the neighborhood of a sample.
In modern applications of interest, samples typically lie near a smaller dimensional manifold although the original space might be high-dimensional. The lower dimensional structures create boundaries, violating the assumptions of existing theoretical analyses where boundary biases might prevail.

Several recent works~\cite{GVG14,GVG15,GOV16} try to resolve the boundary biases for estimating {\em Shannon entropy}. In~\cite{GVG14}, a local SVD was used to enhance the accuracy of the density estimate at sample points. In~\cite{GVG15}, a local Gaussian density with empirical parameters was used to estimate density at sample points. In~\cite{GOV16}, a {\em local likelihood} density estimator (LLDE) was used as the density estimator. Local likelihood density estimator, introduced by~\cite{Loa96,HJ96}, is a systematic approach to resolve the boundary biased of density estimates with mathematically guarantee.
Theoretically, 
the Shannon entropy estimators based on LLDE are known to be consistent,
and empirically they outperform competing estimators under distributions where boundary biases are dominant.

In this paper, we propose to use the local likelihood density estimator as a subroutine and propose an estimator of the integral $J_{\alpha}(X)$, as well as the R\'{e}nyi entropy, based on resubstitution estimators.
%We choose a sample dependent bandwidth (the $k$-NN distance for some fixed $k$). Since the bandwidth is data dependent, the estimator has an asymptotic {\em multiplicative} bias, which can be computed beforehand. We prove that with the correction of the bias, the proposed estimator is $L_2$ consistent. By several synthetic simulation, we demonstrate the advantage of proposed estimator compared to state-of-the-art estimators~\cite{Joe89,Leo08,Pal10}.
The rest of the paper is organized as follows.

\begin{itemize}
    \item In Section~\ref{sec:kde}, we briefly review the kernel density estimator (KDE). We show that KDE with fixed bandwidth suffers from multi-scale data and propose a sample dependent bandwidth which adapts to {\em multi-scale} data. We substitute the KDE with sample dependent bandwidth choice in the resubstitution estimator 
    of $J_{\alpha}(X)$, as well as the R\'{e}nyi entropy. We prove that with the correction of the multiplicative bias, the resulting estimator is $L_2$ consistent.
    \item Even with the local and adaptive choice of the bandwidth,  KDE still suffers from boundary biases.  In Section~\ref{sec:klnn},  we introduce the local likelihood density estimator (LLDE) which can reduce boundary biases, compared to KDE. Again, we establish the $k$-local nearest neighbor ($k$-LNN) estimator of $J_{\alpha}(X)$ based on LLDE and prove its $L_2$ consistency.
    \item In Section~\ref{sec:simul}, we  simulate several synthetic scenarios and compare  the performance of our proposed $k$-LNN estimator, against the traditional KDE based estimators and $k$-NN based estimators~\cite{Leo08}.
\end{itemize}

% ---------------------------------------------------------------------------------------------------------------------------------
\section{Kernel Density Estimator with $k$-NN Bandwidth}
\label{sec:kde}
%\subsection{Kernel Density Estimator with Locally Adaptive Bandwidth}
%We start at the standard {\em kernel density estimator} (KDE).
%Given any point $x$, where we want to estimate the density $f(x)$
%from i.i.d. samples $\{X_1, X_2, \dots, X_n\}$, KDE is given by the following formula:
Given $n$ i.i.d.\ samples $\{X_1, X_2, \dots, X_n\}$ drawn from a distribution $f_X(x)$,
standard {\em Kernel Density Estimator} (KDE) is defined for
a bandwidth $h\in\reals$  and a kernel function
$K: \mathbb{R}^d \to \mathbb{R}^+$ that integrates to 1 as
\begin{eqnarray}
\label{eq:KDE}
\hf_n^{{\rm (KDE)}}(x) &=& \frac{1}{nh^d} \sum_{i=1}^n K\left( \frac{X_i - x}{h} \right) \;.
\end{eqnarray}
%here $K: \mathbb{R}^d \to \mathbb{R}^+$ is the kernel function that integrates to 1 and $h$ is the bandwidth of the kernel.
Typical choices of $K$ include Gaussian kernel $K(u) \propto \exp\{-\|u\|^2/2\}$, uniform kernel $K(u) \propto \mathbb{I}\{\|u\| \leq 1\}$ and Epanechnikov kernel $K(u) \propto (1-\|u\|^2) \mathbb{I}\{\|u\| \leq 1\}$.
% The bandwidth $h$ trades off the bias and variance of the estimator.
The consistency of KDE is known for global choices of $h$ (that does not change for different points $x$)
in the range of $h \to 0$ and $nh^d \to \infty$ as the number of samples $n$ goes to infinity \cite{Was06}.

\begin{figure}[h]
	\begin{center}
	\includegraphics[width=.65\textwidth]{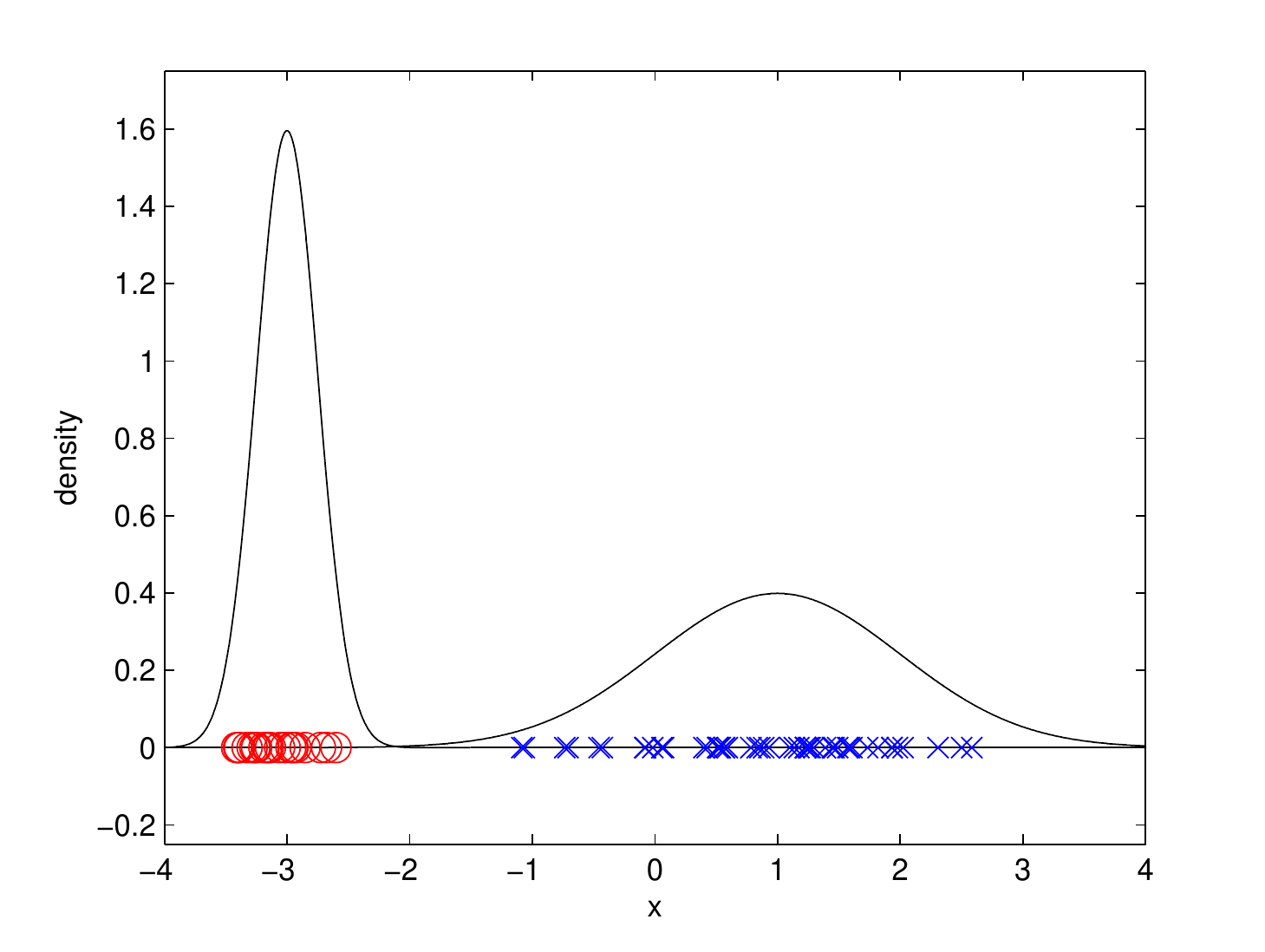}
	%\put(-240,80){\small$ \E[\hJ_2^{(n)}(X)] $}
	%\put(-170,-5){ $(1-r)$ where $r$ is correlation}
	\end{center}
	\caption{An example of
	samples denoted by `x' for one of the mixtures and `o' for the other mixture
	under a mixture of two Gaussians.
	The pdf is shown in a solid black line.
	Fixed bandwidth do not work well for both `x' samples  and `o' samples .}
	\label{fig:fig1}
\end{figure}

%{\bf[Are we the first to use k-NN bandwidth with KDE? Is consistency only known for fixed global bandwidth? I would think variable local bandwidth that scales as $h \to 0$ and $nh^d \to \infty$ would also achieve consistency.]}
Although typical analyses of KDE assume a fixed global bandwidth, in practice there is significant gain in local and variable choice of band widths.
For example, consider a case of  a mixture of two Gaussian distributions  (see Figure~\ref{fig:fig1}).
% then a fixed bandwidth can not work well for every $x$.
A fixed bandwidth choice can be either too large in the  low variance regime of $x$
(labeled by    `o' in Figure~\ref{fig:fig1}) or too small for large variance regime of $x$ (labeled by `x' in Figure~\ref{fig:fig1}).
In real applications in high dimensions, such heterogeneity is prevalent. %Therefore, the fixed bandwidth KDE needs to be improved.

Previous work in~\cite{Rosen56,TW92} suggests using a locally adaptive bandwidth $h(x)$ which varies with $x$.
One previously suggested choice of $h(x)$ is the distance between $x$ and its $k$-th nearest neighbor among $\{X_1, X_2, \dots, X_n\}$.
This choice is referred to as the {\em $k$-NN bandwidth}.
Just as the value of a fixed bandwidth $h$ trades off bias and variance,
now the value of an integer $k$ also  trades off between bias and variance.
We note here that  if the uniform kernel $K(u) \propto \mathbb{I}\{\|u\| \leq 1\}$ combined with $k$-NN bandwidth is used, then KDE reduces to the $k$-NN density estimator.
In~\cite{TW92}, it was shown that if $k$ is a function of $n$ such that $k(n) \to \infty$ and $k(n)/n \to 0$ as $n$ goes to infinity, then the KDE with $k$-NN bandwidth
is consistent. In the example above, the $k$-NN bandwidth adapts to the local geometry of the samples and suffers less from
heterogeneity of data compared to a fixed bandwidth.

In this paper, we propose to use the $k$-NN bandwidth, but with a {\em fixed and small}  $k$ in the range of $4 \sim 8$.
% The reason is that small $k$ will reduce the asymptotic bias of the estimator.
Such a choice, violating  $k \to \infty$, % is required to make the
results in an inconsistent density estimator. %KDE consistent,
However, we propose pre-computing this {\em universal} asymptotic bias and de-biasing the resulting estimator. 
Precisely, we prove that if we plug the KDE with $k$-NN bandwidth into the resubstitution estimator of $J_{\alpha}(X)$, there will be a {\em multiplicative} bias which is {\em independent} of the underlying distribution, and hence can be precomputed and divided off from our estimate.

\subsection{KDE based Estimator of $J_{\alpha}(X)$}

As $J_{\alpha}(X) = \E \left[\, f^{\alpha-1}(X) \,\right]$, we propose a resubstitution estimator of the form
\begin{eqnarray}
	\label{eq:resubstitute}
    	\hJ_{\alpha}(x) \; =\;  \frac{1}{n} \sum_{i=1}^n (\hf_n(X_i))^{\alpha-1}\;,
\end{eqnarray}
where
for the density estimate $\hf_n(X_i)$, we propose KDE in \eqref{eq:KDE} with $k$-NN bandwidth $h=\rho_{k,i}$:
%We plug KDE with $k$-NN bandwith into the resubstitution estimator and obtain the following estimator for $J_{\alpha}(X)$:
\begin{eqnarray}
	\label{eq:def_KDE}
	\hJ_{\alpha}^{{\rm (KDE)}}(X) &=& \frac{1}{n } \sum_{i=1}^n  \frac{1}{B_{k,d,\alpha,K}}\left( \frac{1}{n\rho_{k,i}^d} \sum_{j \in \cS_{i,m}} K\left(\, \frac{X_j - X_i}{\rho_{k,i}}\,\right)\right)^{\alpha-1} \;,
\end{eqnarray}
where $\rho_{k,i}$ is the distance to the $k$-th nearest neighbor from sample $X_i$.
Notice the extra multiplicative factor of  $1/B_{k,d,\alpha,K}$.
This is the de-biasing term that cancels the multiplicative asymptotic bias that is
present in the simple resubstitution estimate that directly substitutes \eqref{eq:KDE} in \eqref{eq:resubstitute}.
We show in the following theorem that the multiplicative bias  $B_{k,d,\alpha,K}$ only depends on
 $k$, $d$, $\alpha$ and the choice of kernel $K$, and not on the underlying distribution $f_X(x)$.
 Hence, it can be pre-computed and divided off as explicitly written in  \eqref{eq:def_KDE}.

% $\rho_{k,i}$ is defined as the distance between $X_i$ and its $k$ nearest neighbor among $\{X_j\}_{j \neq i}$.
In the summation in \eqref{eq:def_KDE},  %To compute the KDE,
we only use the subset of $m = \lceil \log n \rceil$ nearest samples defined as $\cS_{i,m}=\{j\in[n]\,:\, j\neq i \text{ and } \|X_i-X_j\|\leq  \rho_{\lceil\log n \rceil,i} \}$. Such a truncation makes the estimator computationally more efficient, as well as
allows us to provide a sharp analysis on the asymptotic bias.
If we want to include more samples in the computation,
our analysis technique can immediately be generalized
 as long as $m=O(n^{{1/(2d)}-\varepsilon})$ for an arbitrarily small $\varepsilon > 0$.
However, for a larger choice of $m$ such as $m = \Omega(n)$, those sample points that are
further away have
statistical properties that are significantly different from those that are closer, which
requires new analysis techniques.
The following shows that the asymptotic multiplicative bias $B_{k,d,\alpha,K}$ does not depend on
the underlying $f_X(x)$, and hence can be computed beforehand and removed.
\begin{thm}
\label{thm:unbiased_KDE}
Let $X_1, X_2, \dots, X_n \in \mathbb{R}^d$ are i.i.d.\ samples from a twice continuously differentiable pdf $f(x)$ such that $\E \left[\, |f(X)|^{\alpha-1}\,\right] < +\infty$, and $K(u)$ is a kernel function such that $K(u) \leq C \|u\|^{-2d}$ for some constant $C > 0$, then
\begin{eqnarray}
\lim_{n \to \infty} \E [\hJ^{{\rm (KDE)}}_{\alpha}(X)  ]  &=&  J_{\alpha}(X) \;,
\end{eqnarray}

Further, if $\E\left[\, |f(X)|^{2\alpha-2} \,\right]<+\infty$, then the variance of the proposed estimator is bounded by
\begin{eqnarray}
{\rm Var} [\hJ^{{\rm (KDE)}}_{\alpha}(X) ]  &=&  O\Big( \frac{(\log n)^2}{n} \Big) \;.
\end{eqnarray}
\end{thm}

This theorem shows the $L_1$ and $L_2$ consistency of the KDE based estimator of $J_{\alpha}(X)$. Conditional on $X_i = x$,
the estimator is a function of the nearest neighbor statistics $Z_{\ell,i} = X^{(\ell)}_i - x$, where $X^{(\ell)}_i$ is the
$\ell$-nearest neighbor from $x$. The key technical step of the proof is to make a connection between
the  nearest neighbor
statistics and uniform order statistics, shown in Lemma \ref{lem:order_stat}. It is shown that the distances
$\rho_{\ell,1} = \|Z_{\ell,1}\|$'s jointly converge to the standardized uniform order statistics, and the directions
$(X_{j_\ell} -X_i)/\|X_{j_\ell} -X_i\|$'s converge to i.i.d.\ random variables drawn uniformly over the unit sphere
in $\reals^d$ (which is called the Haar random variable), jointly with the distances as well.

\begin{lemma}[Lemma 3.2.~\cite{GOV16}]
    \label{lem:order_stat}
    Let $E_1, E_2, \dots, E_m$ be i.i.d.\ standard exponential random variables and
    $\xi_1 , \xi_2, \dots, \xi_m$ be i.i.d.\ random variables  drawn uniformly over the unit $(d-1)$-dimensional sphere
    in $d$ dimensions,
    independent of the $E_i$'s.
    Suppose $f$ is twice continuously differentiable and $x \in \mathbb{R}^d$ satisfies that there exists $\varepsilon > 0$ such that $f(a) > 0$, $\|\nabla f(a)\| = O(1)$ and $\|H_f(a)\| = O(1)$ for any $\|a - x\| < \varepsilon$.
    Then for any $m = O( \log n)$,
    we have the following convergence  conditioned on  $X_i = x$:
    \begin{eqnarray}
    \lim_{n \to \infty} d_{\rm TV} ( (c_d nf(x))^{1/d} (\, Z_{1,i},  \dots, Z_{m,i} \,) \;,\;
    (\, \xi_1 E_1^{1/d}, \dots , \xi_{m}(\sum_{\ell=1}^{m} E_\ell)^{1/d} \,) ) = 0 \;.
    \end{eqnarray}
    where $d_{\rm TV}(\cdot,\cdot)$ is the total variation and $c_d$ is the volume of unit Euclidean ball in $\mathbb{R}^d$.%$\|X - Y\|_{TV} = \sup_B |\,\Pr\{X \in B\} - \Pr\{Y \in B\}\,|$ is the total variation distance of two random variables $X$ and $Y$.
\end{lemma}

Given Lemma~\ref{lem:order_stat}, we  show that the quantity $S = \sum_{j \in \cS_{i,m}} K\left(\, (X_j - X_i)/\rho_{k,i}\,\right)$
 used in the estimate~\eqref{eq:def_KDE} converges in distribution,
 and we can characterize the asymptotic distribution exactly using uniform order statistics.
% similar quantities defined by order statistics as $n$ tends to infinity.
 For  i.i.d.\ standard exponential random variables $E_1,E_2,\ldots,E_m$  and i.i.d.\ Haar random variables
 $\xi_1,\xi_2,\ldots,\xi_m$ in $\reals^d$, we define,
\begin{eqnarray}
	\tilde{S}^{(m)} \equiv  \sum_{j=1}^{m} K\left(\, \frac{\xi_j (\sum_{\ell=1}^j E_{\ell})^{1/d}}{(\sum_{\ell=1}^k E_{\ell})^{1/d}}\,\right) \label{eq:S}\;,
\end{eqnarray}
and let $\tS = \lim_{m\to \infty} \tS^{(m)}$. We can show that if the kernel satisfies $K(u) \leq C \|u\|^{-2d}$ (which is fulfilled by all kernels with bounded support or exponentially decaying tails), the limit of $\tS$ exists and is related to the multiplicative bias term $B_{k,d,\alpha,K}$ in the resubstitution estimator of $J_{\alpha}(X)$ in \eqref{eq:def_KDE}:
% When $d=1$, $\xi_j$'s are  Rademacher random variables.
\begin{eqnarray}
	B_{k,d,\alpha,K} = \E\left[\, \left( \frac{c_d \tS}{\sum_{\ell=1}^k E_\ell} \right)^{\alpha-1} \,\right] \;.
	\label{eq:defBias_KDE}
\end{eqnarray}
where $c_d$ is the volume of the unit ball in $\reals^d$. We provide a proof in Section~\ref{sec:proof_KDE}.
Below is a table of $B_{k,d,\alpha, K}$ computed via numerical simulations, for the Gaussian kernel $K \propto \exp\{-\|u\|^2/2\}$ and some typical values of $k$, $d$ and $\alpha$.
Here $1.0245(\pm 3)$ means the bias has empirical mean $\mu=10245\times 10^{-4}$ with confidence interval $3\times10^{-4}$. We run 1,000,000 trials with
truncation of the summation at $m=5,000$ in these simulations.

\begin{table}[h]
\begin{center}
  \begin{tabular}{  c c | c | c | c | c | c | c   |}
	\cline{3-8}
    & & \multicolumn{6}{|c|}{$k$} \\ \cline{3-8}
     & &  $4$ & $5$ & $6$ & $7$ & $8$ & $9$ \\ \cline{3-8}   \hline
     \multicolumn{1}{ |c  }{\multirow{2}{*}{$d=1$} } &
     \multicolumn{1}{ |c|| }{$\alpha=2$} & $1.0245(\pm 3)$ & $1.0184(\pm 3)$ & $1.0153(\pm 3)$ & $1.0132(\pm 2)$ & $1.0114(\pm 2)$ & $1.0098(\pm 2)$
         \\ \cline{2-8}
     \multicolumn{1}{ |c  }{} &
     \multicolumn{1}{ |c|| }{$\alpha=3$} & $1.1973(\pm 8)$ & $1.1564(\pm 7)$ & $1.1282(\pm 6)$ & $1.1078(\pm 6)$ & $1.0945(\pm 5)$ & $1.10835(\pm 5)$
          \\ \cline{1-8}    \hline
     \multicolumn{1}{ |c  }{\multirow{2}{*}{$d=2$} } &
     \multicolumn{1}{ |c|| }{$\alpha=2$} & $0.9883(\pm 2)$ & $0.9897(\pm 2)$ & $0.9915(\pm 2)$ & $0.9930(\pm 1)$ & $0.9934(\pm 1)$ & $0.9943(\pm 1)$
         \\ \cline{2-8}
     \multicolumn{1}{ |c  }{} &
     \multicolumn{1}{ |c|| }{$\alpha=3$} & $1.0431(\pm 5)$ & $1.0342(\pm 4)$ & $1.0270(\pm 4)$ & $1.0226(\pm 3)$ & $1.0196(\pm 3)$ & $1.0175(\pm 3)$
          \\ \cline{1-8}    \hline
     \multicolumn{1}{ |c  }{\multirow{2}{*}{$d=3$} } &
     \multicolumn{1}{ |c|| }{$\alpha=2$} & $0.9821(\pm 1)$ & $0.9856(\pm 1)$ & $0.9883(\pm 1)$ & $ 0.9900(\pm 1)$ & $0.9912(\pm 1)$ & $0.9920(\pm 1)$
         \\ \cline{2-8}
     \multicolumn{1}{ |c  }{} &
     \multicolumn{1}{ |c|| }{$\alpha=3$} & $0.9926(\pm 3)$ & $0.9935(\pm 2)$ & $0.9940(\pm 2)$ & $0.9954(\pm 2)$ & $0.9951(\pm 2)$ & $0.9955(\pm 2)$
          \\ \cline{1-8}
  \end{tabular}
\end{center}
     \caption{Numerical approximation of $B_{k,d,\alpha,K}$ for the Gaussian kernel.}
     \label{tbl:bias_KDE}\end{table}

\subsection{KDE based R\'{e}nyi entropy estimator}

Given the KDE based estimator for $J_{\alpha}(X)$, we propose the following estimator for the R\'{e}nyi entropy,
\begin{eqnarray}
&& \widehat{H}^{{\rm (KDE)}}_{\alpha}(X) = \frac{1}{1-\alpha} \log \widehat{J}^{{\rm (KDE)}}_{\alpha}(X) \,\notag\\
&=& \frac{1}{1-\alpha} \left(\, \log \sum_{i=1}^n  \left( \frac{1}{n\rho_{k,i}^d} \sum_{j \in \cS_{i,m}} K\left(\, \frac{X_j - X_i}{\rho_{k,i}}\,\right)\right)^{\alpha-1} - \log n - \log B_{k,d,\alpha,K} \,\right) \;.
\end{eqnarray}

Following by the $L_2$ consistency of $\hJ^{{\rm (KDE)}}_{\alpha}(X)$ and the fact that $\log(\cdot)$ is continuous on $\mathbb{R}^+$, we obtain the following corollary showing convergence property of $\hH^{{\rm (KDE)}}_{\alpha}(X)$.
\begin{coro}
    \label{cor:renyi_unbiased_KDE}
    Under the same assumption of Theorem~\ref{thm:unbiased_KDE}, the estimator $\hH^{{\rm (KDE)}}_{\alpha}(X)$ converges to $H_{\alpha}(X)$ in probability, as $n \to \infty$.
\end{coro}

%---------------------------------------
\section{Local Likelihood Density Estimator with $k$-NN Bandwidth}
\label{sec:klnn}

%\subsection{Local Likelihood Density Estimator}
In this section, we propose the {\em local likelihood density estimator} (LLDE), introduced in~\cite{Loa96,HJ96}, as a generalization of KDE. 
In practice, the choice of a bandwidth is mostly left to the practitioner -- here we propose using the $k$-NN bandwidth for LLDE. Given a point $x$ and i.i.d.\ samples $\{X_1, X_2, \dots, X_n\}$, the LLDE is given by \cite{loader2006local,GOV16}:
\begin{eqnarray}
\label{eq:LLDE}
\widehat{f}_n^{{\rm (LLDE)}}(x) &\equiv& \frac{S_0}{n(2\pi)^{d/2} h^d |\Sigma|^{1/2}} \exp\{-\frac{1}{2}\mu^T \Sigma^{-1} \mu\} \;,
\end{eqnarray}
where the quantities $S_0$, $S_1$, $S_2$ and $\mu$, $\Sigma$ are defined as follows,
\begin{eqnarray}
S_0 &\equiv& \sum_{j=1}^n e^{-\frac{\|X_j - x\|^2}{2 h^2}} \;, \label{eq:defS0}\\
S_1 &\equiv& \sum_{j=1}^n \frac{X_j - x}{\rho_{k,i}} \, e^{-\frac{\|X_j - x\|^2}{2 h^2}} \;,\label{eq:defS1}\\
S_2 &\equiv& \sum_{j=1}^n \frac{(X_j - x)(X_j - x)^T}{\rho_{k,i}^2} \, e^{-\frac{-|X_j - x\|^2}{2 h^2}} \;,\label{eq:defS2}\\
\mu &\equiv& \frac{S_1}{S_0} \;,\label{eq:defmu}\\
\Sigma &\equiv& \frac{S_2}{S_0} - \frac{S_1 S_1^T}{S_0^2} \;,\label{eq:defSigma}
\end{eqnarray}
and for the bandwidth, we propose  using the $k$-NN distance: $h=\rho_{k,i}$.

LLDE can be viewed as a weighted local Gaussian density, where the Gaussian kernel $K((X_j -x)/h) \propto \exp\{-\|X_j-x\|/(2 h^2)\}$ is used to compute the weight from samples.
Locally, in the neighborhood of a sample point $X_i$, $\mu = S_1/S_0$ is the weighted sample mean and $\Sigma = S_2/S_0 - S_1S_1^T/S_0^2$ is the weighted sample variance. Notice that the KDE estimator with $k$-NN bandwidth at point $x$ can be written as $\hf_n^{{\rm (KDE)}}(x) = S_0/(n(2\pi)^{d/2}h^d)$.
Compared with LLDE, KDE can be viewed as a weighted local Gaussian density where the mean is restricted to be $x$ and the variance is restricted to be identity. Therefore, LLDE is able to capture the local structure automatically, hence can reduce the boundary bias if $x$ is near the boundary of the density.

\begin{figure}[htbp]
	\begin{center}
	\includegraphics[width=0.45\textwidth]{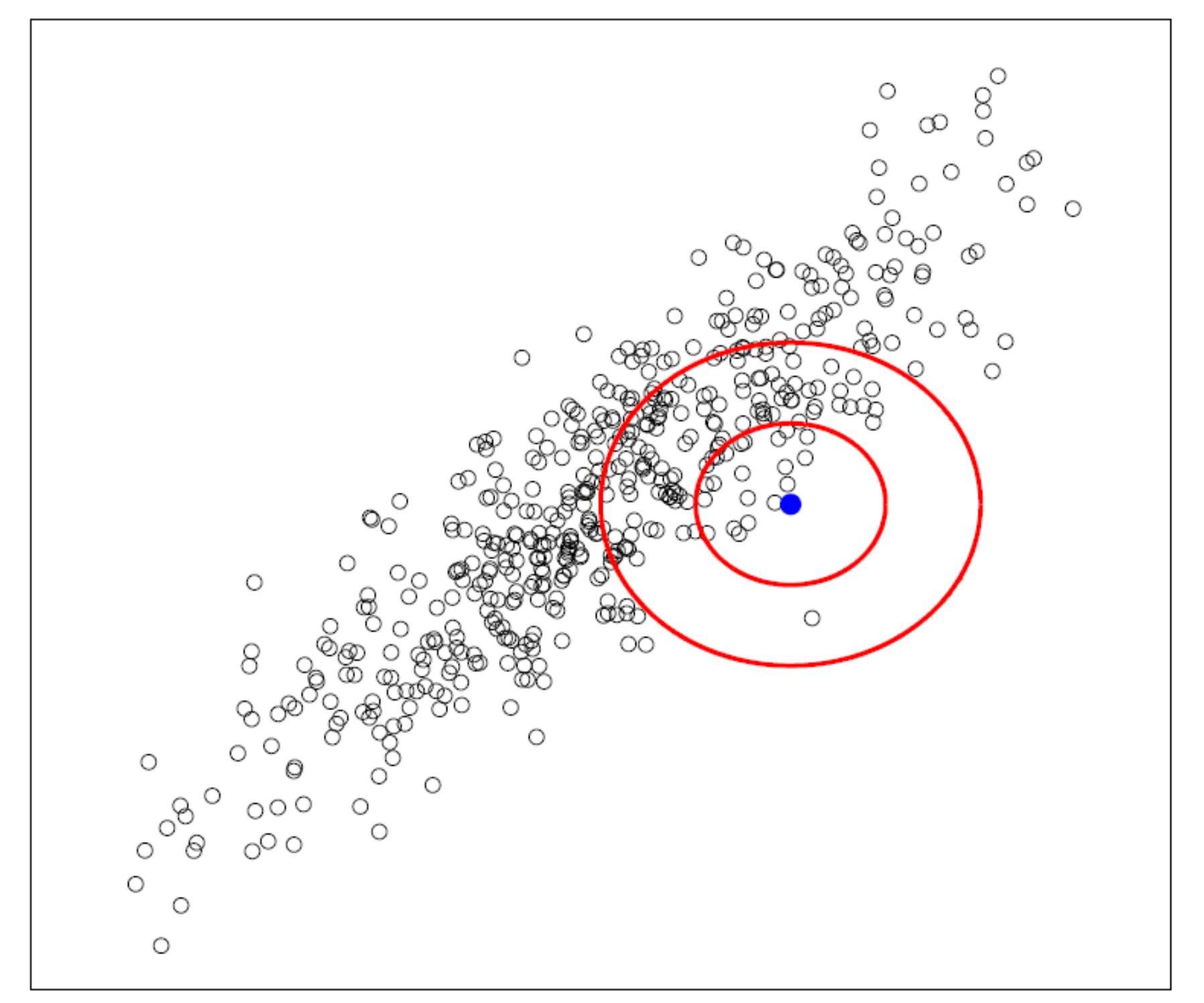}
	\put(-220,80){$ Y $}
	\put(-120,-15){ $X$}
	\hspace{0.9 cm}
	\includegraphics[width=0.45\textwidth]{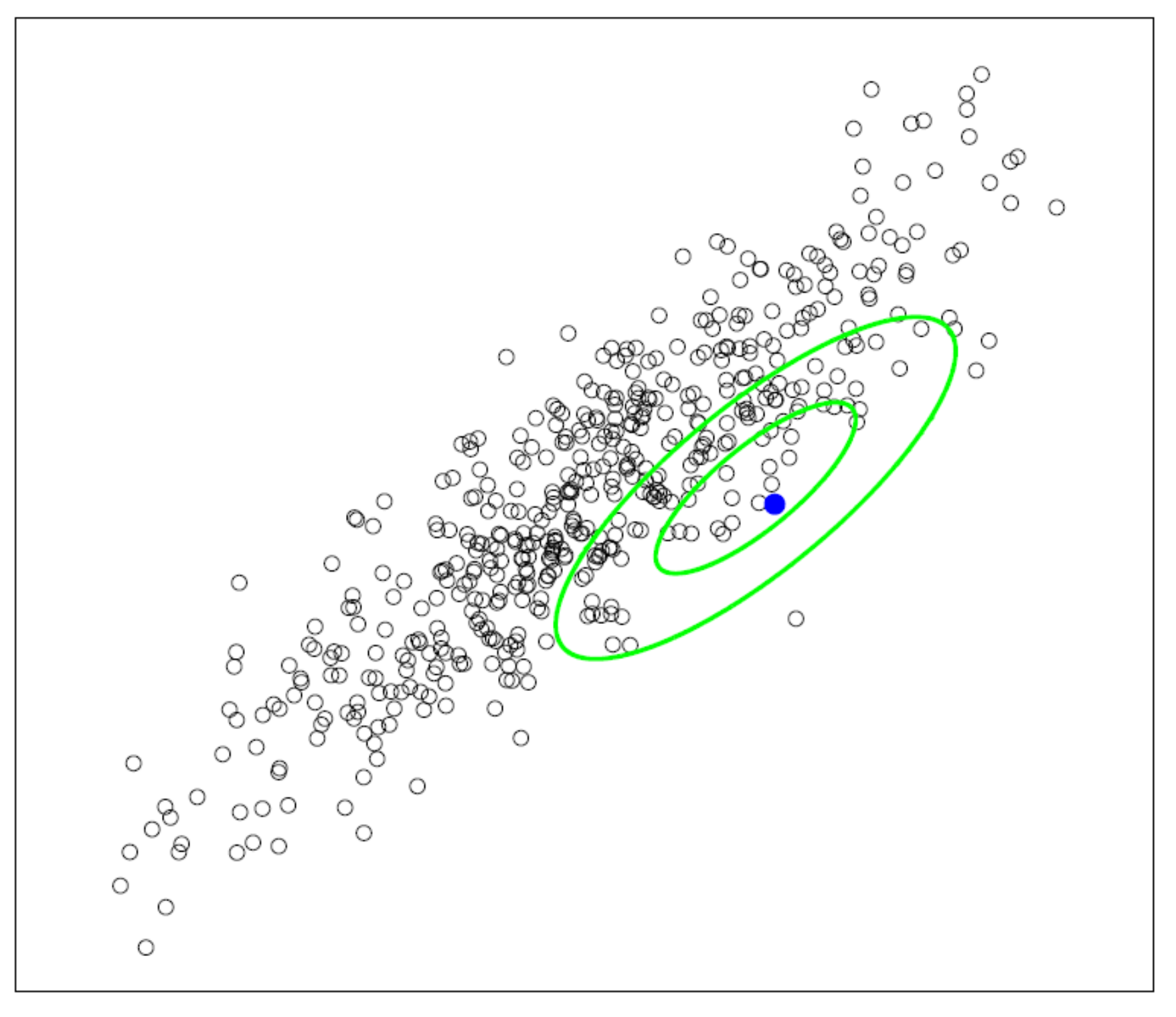}
	\put(-225,80){$ Y$}
	\put(-120,-15){$X$}
	\end{center}
	\caption{Given samples from joint Gaussian distribution as an example,
	consider approximating the local density near  the blue point $x$ near boundary of the distribution,
	using a Gaussian density with mean $x$ and unit variance (left) and a Gaussian density with local sample mean and covariance (right).}
	\label{fig_2}
\end{figure}

In Figure~\ref{fig_2}, the data are drawn from highly correlated joint Gaussian distribution, where we want to estimate the density of the blue point $x$ near the boundary. On the left, the red contours show that KDE is a Gaussian density with mean $x$ and unit variance, while on the right the green contours corresponds to a Gaussian density with weighted sample mean and variance given by LLDE. We can see that LLDE fits the local structure better than KDE, capturing the fact that $x$ is at the boundary of the underlying density.

\subsection{LLDE based Estimator of $J_{\alpha}(X)$}

We substitute  LLDE in the resubstitution estimator $\hJ_{\alpha}(X) = (1/n) \sum_{i=1}^n (\hf_{\alpha}(X_i))^{\alpha-1}$ to obtain the following $k$-Local Nearest Neighbor ($k$-LNN) estimator of the integral $J_{\alpha}(X)$,
\begin{eqnarray}
    \label{eq:def_kLNN}
    \hJ_{\alpha}^{(k-{\rm LNN})}(x) &=& \frac{1}{n B_{k,d,\alpha}} \sum_{i=1}^n  \left( \frac{S_{0,i}}{n(2\pi)^{d/2}\rho_{k,i}^d |\Sigma_i|^{1/2}} \exp\{-\frac{1}{2} \mu_i^T \Sigma_i^{-1} \mu_i\}\right)^{\alpha-1} \;,
\end{eqnarray}
here $B_{k,d,\alpha}$ is again the multiplicative bias that depends on $k$, $d$ and $\alpha$, but not the underlying distribution.
Recall that  $\rho_{k,i}$ is the distance between $X_i$ and its $k$-th nearest neighbor.
The quantities $S_{0,i}$, $S_{1,i}$, $S_{2,i}$ and $\mu_i$, $\Sigma_i$ are
defined from \eqref{eq:defS0}-\eqref{eq:defSigma} in the neighborhood of a sample point $x=X_i$,
and with a choice of the bandwidth $h=\rho_{k,i}$.
%given as follows,
%\begin{eqnarray}
%\label{eq:defS_i}
%S_{0,i} &\equiv& \sum_{j \in \cS_{i,m}} e^{-\frac{\|X_j - X_i\|^2}{2\rho_{k,i}^2}} \;,\\
%S_{1,i} &\equiv& \sum_{j \in \cS_{i,m}} \frac{X_j - X_i}{\rho_{k,i}} \, e^{-\frac{\|X_j - X_i\|^2}{2\rho_{k,i}^2}} \;,\\
%S_{2,i} &\equiv& \sum_{j \in \cS_{i,m}} \frac{(X_j - X_i)(X_j - X_i)^T}{\rho_{k,i}^2} \, e^{-\frac{-|X_j - X_i\|^2}{2\rho_{k,i}^2}} \;,\\
%\mu_i &\equiv& \frac{S_{1,i}}{S_{0,i}} \;,\\
%\Sigma_i &\equiv& \frac{S_{2,i}}{S_{0,i}} - \frac{S_{1,i} S_{1,i}^T}{S_{0,i}^2} \;.
%\end{eqnarray}
Similar to the KDE based estimator~\eqref{eq:def_KDE}, only the subset of $m = \lceil \log n \rceil$ nearest samples $\cS_{i,m}$ are used for computing the quantities for the same reason. The following theorem shows the $L_1$ and $L_2$ consistency of the $k$-LNN estimator of $J_{\alpha}(X)$ for twice continuously differentiable density $f(x)$.
%{\bf MOVE THE DEFINITION THAT IS COMMENTED OUT TO THE APPENDIX SOMEWHERE.}

\begin{thm}
\label{thm:unbiased_kLNN}
Let $X_1, X_2, \dots, X_n \in \mathbb{R}^d$ are i.i.d.\ samples from a twice continuously differentiable pdf $f(x)$ such that $\E \left[\, |f(X)|^{\alpha-1}\,\right] < +\infty$, then
\begin{eqnarray}
\lim_{n \to \infty} \E [\hJ^{(k-{\rm LNN})}_{\alpha}(X)  ]  &=&  J_{\alpha}(X) \;,
\end{eqnarray}

If $\E\left[\, |f(X)|^{2\alpha-2} \,\right]<+\infty$, then the variance of the proposed estimator is bounded by
\begin{eqnarray}
{\rm Var} [\hJ^{(k-{\rm LNN})}_{\alpha}(X) ]  &=&  O\Big( \frac{(\log n)^2}{n}\Big) \;.
\end{eqnarray}
\end{thm}

The idea of the proof is quite similar to that of Theorem~\ref{thm:unbiased_KDE}. For i.i.d.\ standard exponential random variables $E_1,E_2, \dots, E_m$ and i.i.d.\ Haar random variables $\xi_1, \dots, \xi_m$, we define for $\gamma \in \{0,1,2\}$,
\begin{eqnarray}
\tS_{\gamma}^{(m)} = \sum_{j=1}^m \xi_j^{(m)} \frac{(\sum_{\ell=1}^j E_{\ell})^{\gamma}}{(\sum_{\ell=1}^k E_{\ell})^{\gamma}} \exp\{-\frac{(\sum_{\ell=1}^j E_{\ell})^2}{2(\sum_{\ell=1}^k E_{\ell})^2}\} \;,
\end{eqnarray}
where $\xi_j^{(0)} = 1$, $\xi_j^{(1)} = \xi_j \in \mathbb{R}^d$ and $\xi_j^{(2)} = \xi_j \xi_j^T \in \mathbb{R}^{d \times d}$, and $\tS_{\gamma} = \lim_{m \to \infty} \tS_{\gamma}^{(m)}$. $\tilde{\mu} = \tS_1/\tS_0$ and $\tilde{\Sigma} = \tS_2/\tS_0 - \tS_1\tS_1^T/\tS_0^2$.  We show that the quantities $\{S_{0,i},S_{1,i},S_{2,i},\mu_i,\Sigma_i\}$ jointly converge to $\{\tS_0, \tS_1, \tS_2, \tilde{\mu}, \tilde{\Sigma}\}$ using Lemma~\ref{lem:order_stat}. The multiplicative bias $B_{k,d,\alpha}$ is given by,

\begin{eqnarray}
	B_{k,d,\alpha} = \E\left[\, \left( \frac{c_d \tS_0}{(\sum_{\ell=1}^k E_{\ell})(2\pi)^{d/2}|\tilde{\Sigma}|^{1/2}} \exp\{-\frac{1}{2} \tilde{\mu}^T \tilde{\Sigma}^{-1} \tilde{\mu}\} \right)^{\alpha-1} \,\right] \;.
	\label{eq:defBias_kLNN}
\end{eqnarray}

We provide a proof  in Section~\ref{sec:proof_kLNN}. Here we enumerate the approximate value of $B_{k,d,\alpha}$ for some typical $k$, $d$ and $\alpha$.
%Here $1.104(\pm 5)$ means the bias has empirical mean $\mu=1.104$ with confidence interval $5\times10^{-3}$.
 We run 10,000 trials with truncation of the summation at $m=5,000$ in these simulations.

\begin{table}[h]
\begin{center}
  \begin{tabular}{  c c | c | c | c | c | c | c   |}
	\cline{3-8}
    & & \multicolumn{6}{|c|}{$k$} \\ \cline{3-8}
     & &  $4$ & $5$ & $6$ & $7$ & $8$ & $9$ \\ \cline{3-8}   \hline
     \multicolumn{1}{ |c  }{\multirow{2}{*}{$d=1$} } &
     \multicolumn{1}{ |c|| }{$\alpha=2$} & $1.104(\pm 5)$ & $1.076(\pm 4)$ & $1.062(\pm 4)$ & $1.050(\pm 3)$ & $1.045(\pm 3)$ & $1.037(\pm 3)$
         \\ \cline{2-8}
     \multicolumn{1}{ |c  }{} &
     \multicolumn{1}{ |c|| }{$\alpha=3$} & $1.493(\pm 18)$ & $1.358(\pm 10)$ & $1.273(\pm 9)$ & $1.242(\pm 8)$ & $1.199(\pm 7)$ & $1.180(\pm 7)$
          \\ \cline{1-8}    \hline
     \multicolumn{1}{ |c  }{\multirow{2}{*}{$d=2$} } &
     \multicolumn{1}{ |c|| }{$\alpha=2$} & $1.006(\pm 3)$ & $1.003(\pm 3)$ & $1.003(\pm 3)$ & $1.000(\pm 3)$ & $0.994(\pm 2)$ & $0.996(\pm 2)$
         \\ \cline{2-8}
     \multicolumn{1}{ |c  }{} &
     \multicolumn{1}{ |c|| }{$\alpha=3$} & $1.158(\pm 8)$ & $1.139(\pm 7)$ & $1.095(\pm 6)$ & $1.089(\pm 6)$ & $1.073(\pm 5)$ & $1.075(\pm 5)$
          \\ \cline{1-8}    \hline
     \multicolumn{1}{ |c  }{\multirow{2}{*}{$d=3$} } &
     \multicolumn{1}{ |c|| }{$\alpha=2$} & $0.971(\pm 3)$ & $0.977(\pm 2)$ & $0.975(\pm 2)$ & $ 0.978(\pm 2)$ & $0.984(\pm 2)$ & $0.984(\pm 2)$
         \\ \cline{2-8}
     \multicolumn{1}{ |c  }{} &
     \multicolumn{1}{ |c|| }{$\alpha=3$} & $1.034(\pm 5)$ & $1.026(\pm 5)$ & $1.015(\pm 4)$ & $1.011(\pm 4)$ & $1.008(\pm 4)$ & $1.014(\pm 3)$
          \\ \cline{1-8}
  \end{tabular}
\end{center}
     \caption{Numerical approximation of $B_{k,d,\alpha}$.}
     \label{tbl:bias_kLNN}\end{table}

\subsection{$k$-LNN R\'{e}nyi entropy estimator}

Given the $k$-LNN estimator for $J^{(k-{\rm LNN})}_{\alpha}(X)$, we propose the following estimator for the R\'{e}nyi entropy: 
\begin{eqnarray}
&& \widehat{H}^{(k-{\rm LNN})}_{\alpha}(X) = \frac{1}{1-\alpha} \log \widehat{J}^{(k-{\rm LNN})}_{\alpha}(X) \,\notag\\
&=& \frac{1}{1-\alpha} \left(\, \log \sum_{i=1}^n  \left( \frac{S_{0,i}}{n(2\pi)^{d/2}\rho_{k,i}^d |\Sigma_i|^{1/2} } \exp\{- \frac12 \mu_{i}^T  \Sigma_i^{-1} \mu_{i}\} \right)^{\alpha-1} - \log n - \log B_{k,d,\alpha} \,\right) \;.
\end{eqnarray}

Similar to Corollary~\ref{cor:renyi_unbiased_KDE}, by the $L_2$ consistency of $\hJ^{(k-{\rm LNN})}_{\alpha}(X)$ and the fact that $\log(\cdot)$ is continuous on $\mathbb{R}^+$, the $k$-LNN estimator $\hH^{(k-{\rm LNN})}_{\alpha}(X)$ converges to $H_{\alpha}(X)$ in probability, as $n \to \infty$.

%------------------------------------------------------------------------------------------------------------------------------------
\section{Simulations}
\label{sec:simul}
\begin{figure}[h]
	\begin{center}
	\includegraphics[width=.45\textwidth]{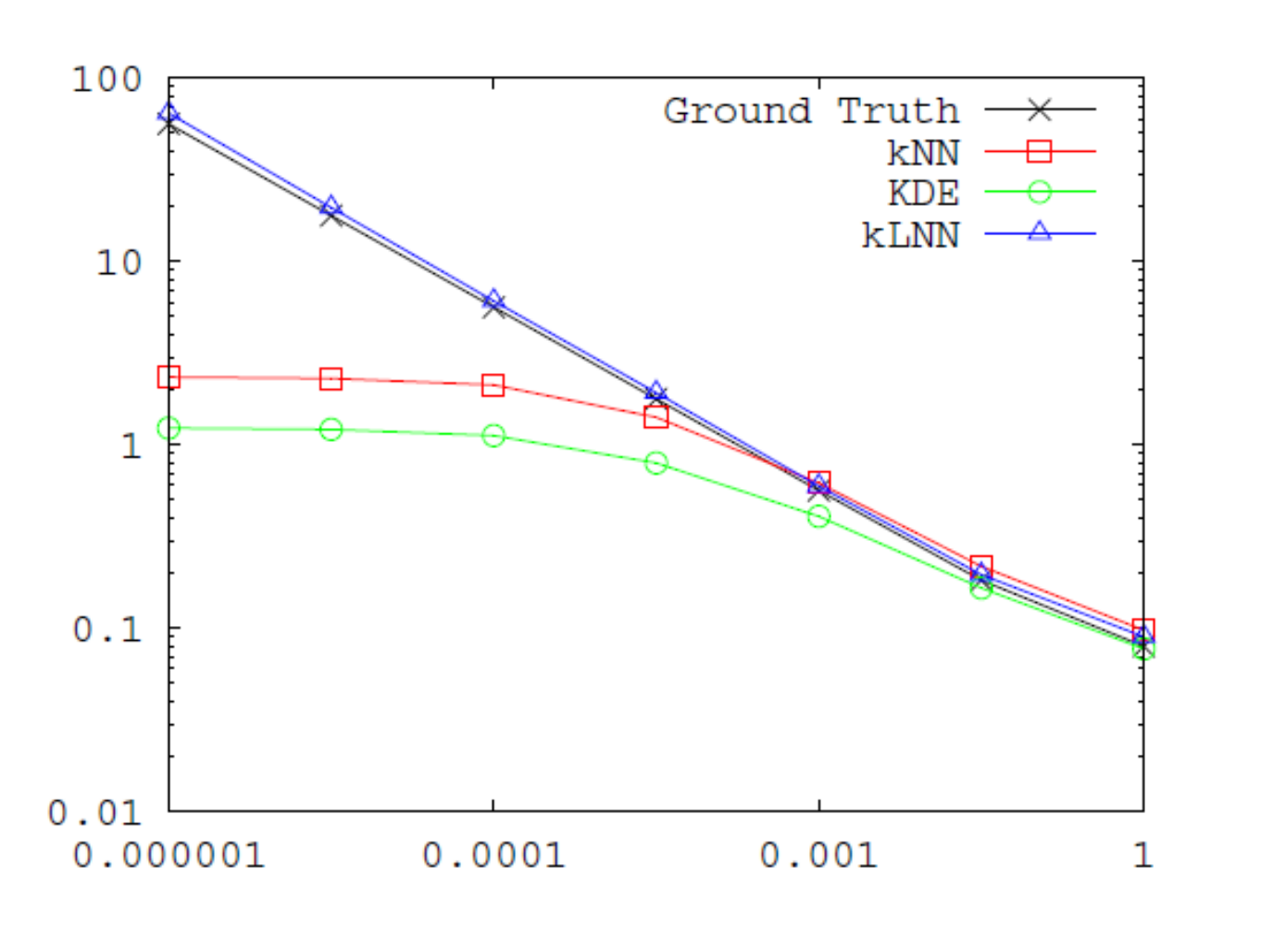}
	\put(-240,80){\small$ \E[\hJ_2(X)] $}
	\put(-170,-0){ $(1-r)$ where $r$ is correlation}
	\hspace{0.9 cm}
	\includegraphics[width=.45\textwidth]{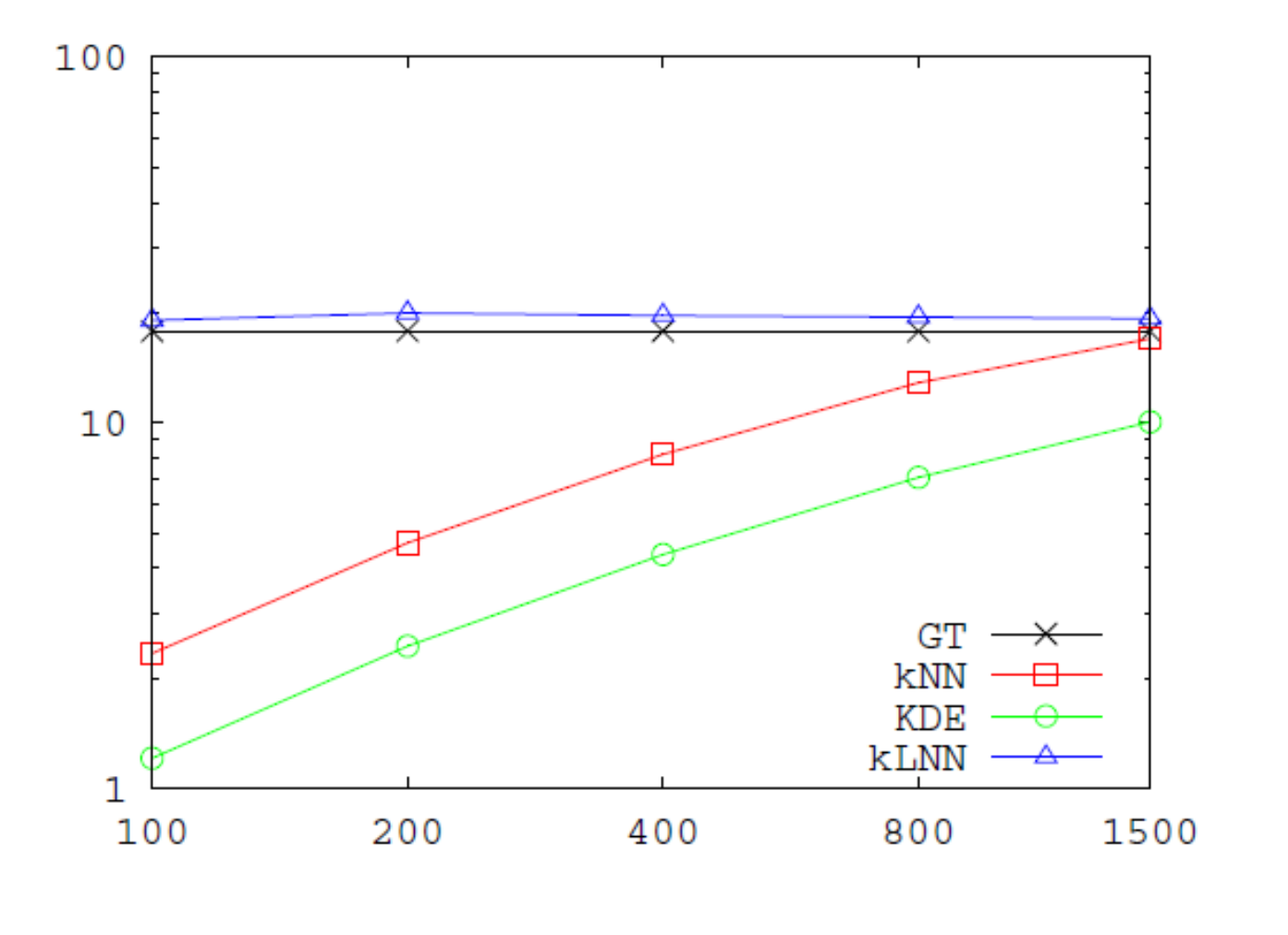}
	\put(-245,80){\small$ \E[\hJ_2(X)] $}
	\put(-150,-0){ number of samples $n$}
	\end{center}
	\caption{Proposed estimator outperform other estimators for $J_2(X)$ for highly correlated Gaussian.}
	\label{fig:d=1_alpha=2}
\end{figure}

\begin{figure}[h]
	\begin{center}
	\includegraphics[width=.45\textwidth]{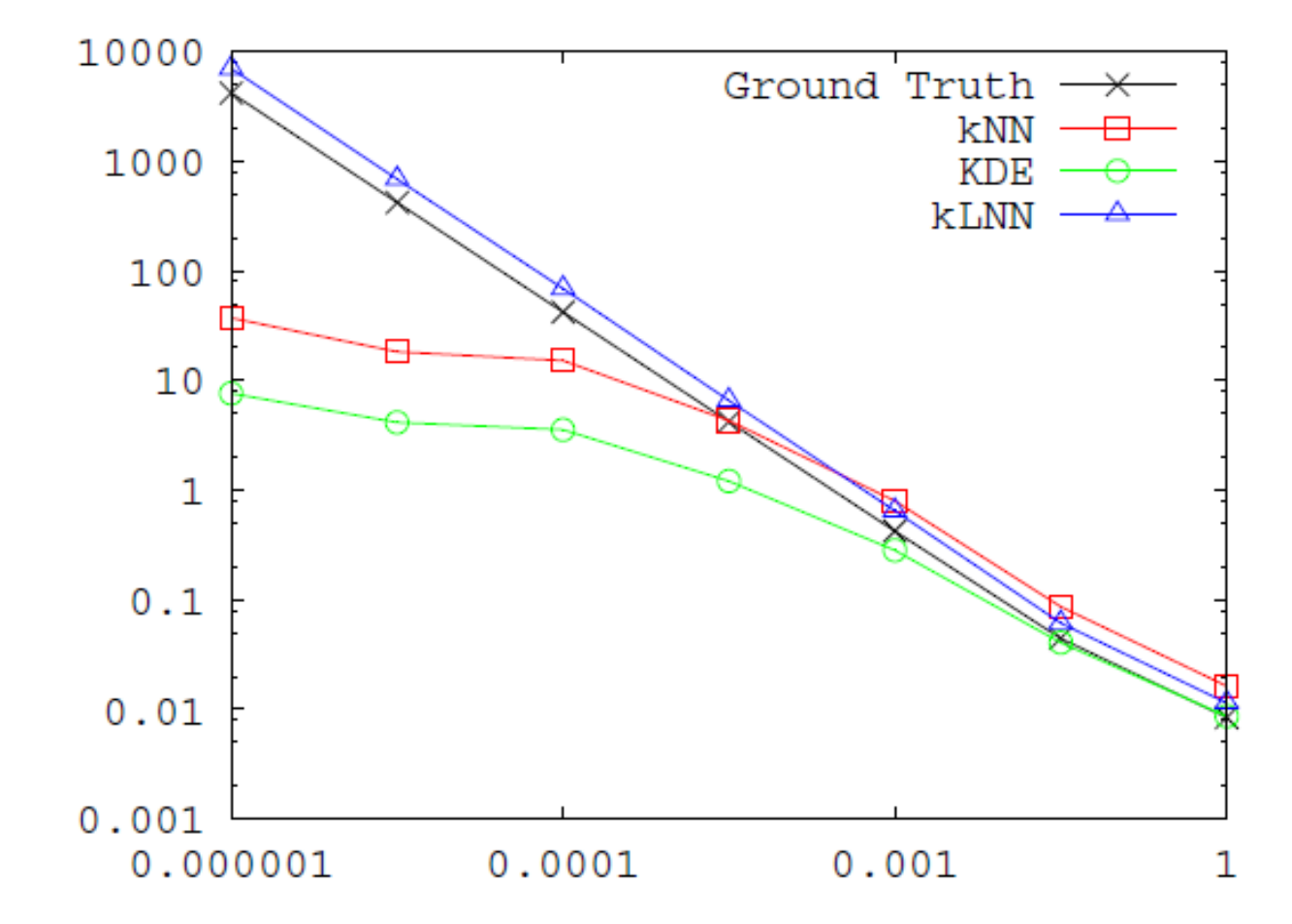}
	\put(-240,80){\small$ \E[\hJ_3(X)] $}
	\put(-170,-5){ $(1-r)$ where $r$ is correlation}
	\hspace{0.9 cm}
	\includegraphics[width=.45\textwidth]{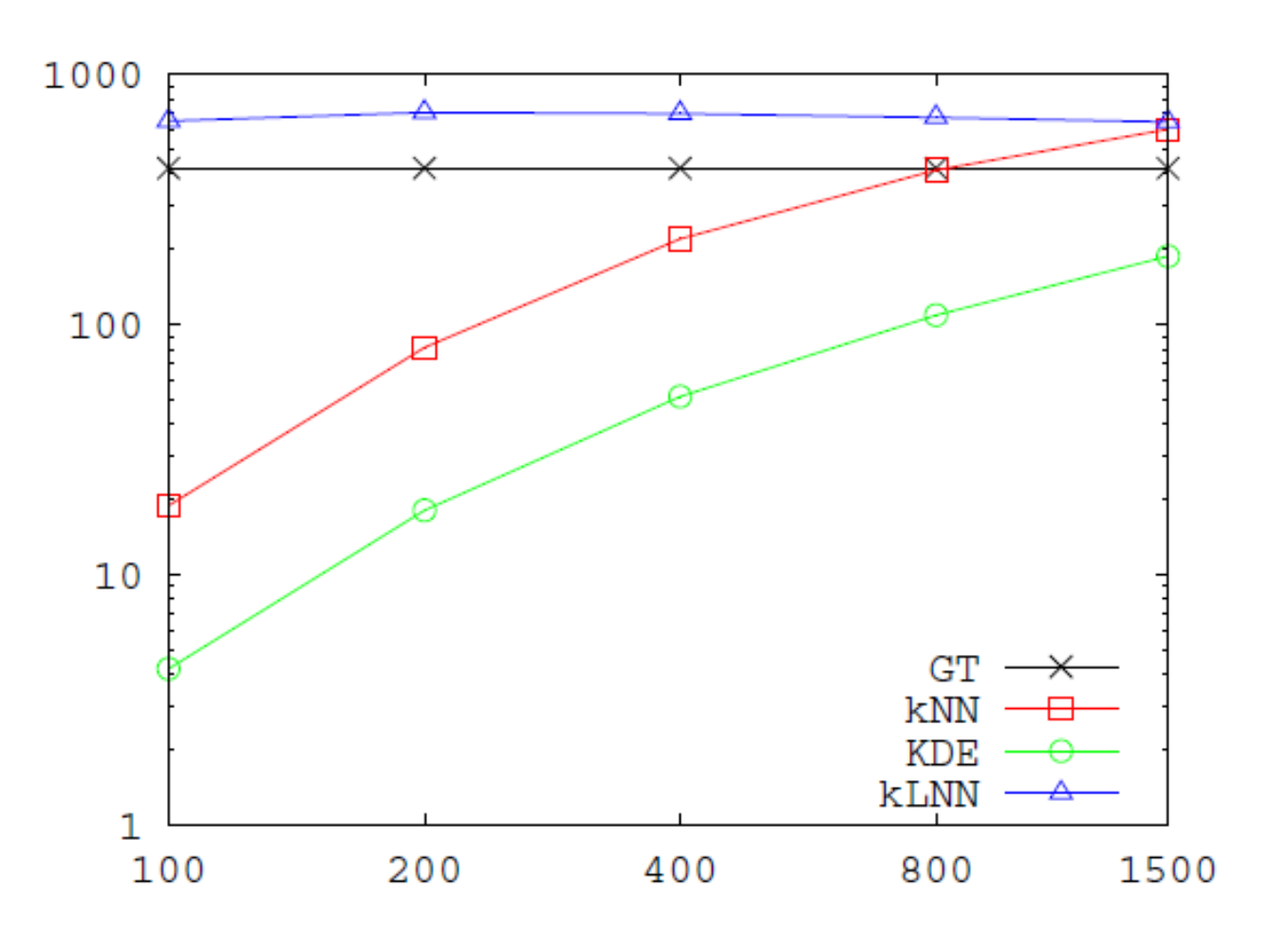}
	\put(-240,80){\small$ \E[\hJ_3(X)] $}
	\put(-150,-0){ number of samples $n$}
	\end{center}
	\caption{Proposed estimator outperform other estimators for $J_3(X)$ for highly correlated Gaussian.}
	\label{fig:d=1_alpha=3}
\end{figure}

\begin{figure}[h]
	\begin{center}
	\includegraphics[width=.45\textwidth]{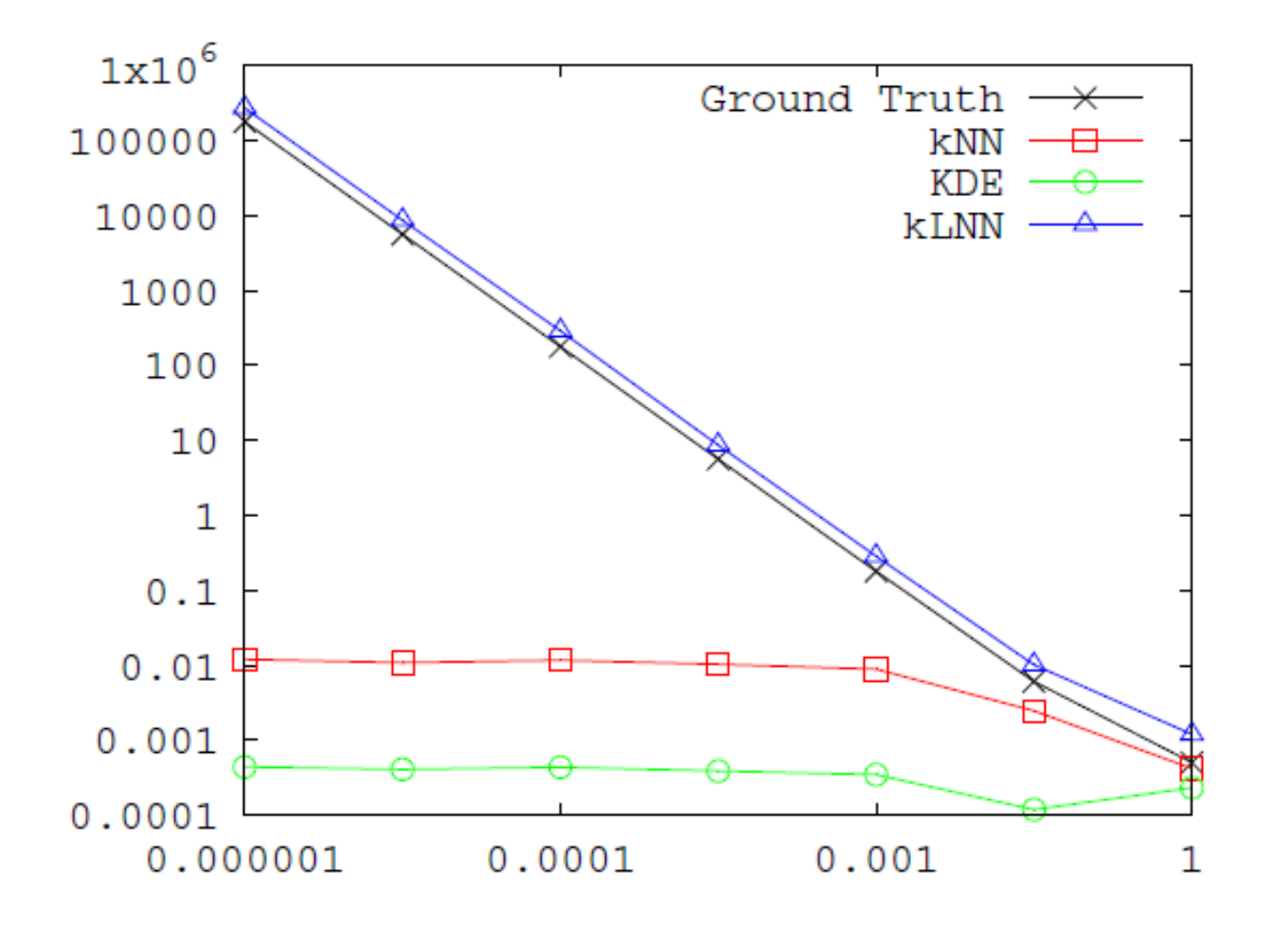}
	\put(-235,80){\small$ \E[\hJ_2(X)] $}
	\put(-170,-0){ $(1-r)$ where $r$ is correlation}
	\hspace{0.9 cm}
	\includegraphics[width=.45\textwidth]{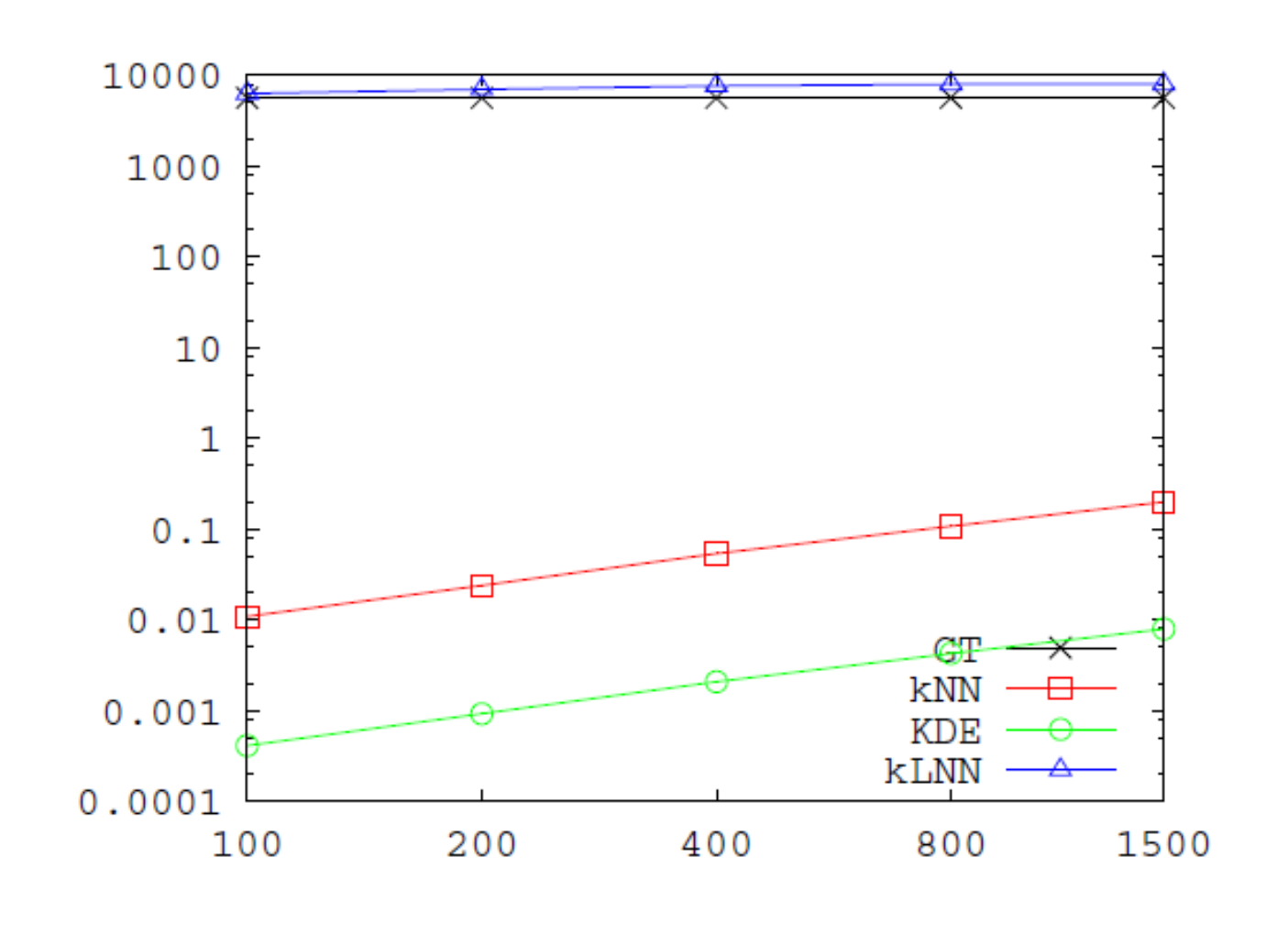}
	\put(-235,80){\small$ \E[\hJ_2(X)] $}
	\put(-150,-0){ number of samples $n$}
	\end{center}
	\caption{Proposed estimator outperform other estimators for $J_2(X)$ for high-dimensional highly correlated Gaussian.}
	\label{fig:d=3_alpha=2}
\end{figure}

\begin{figure}[h]
	\begin{center}
	\includegraphics[width=.45\textwidth]{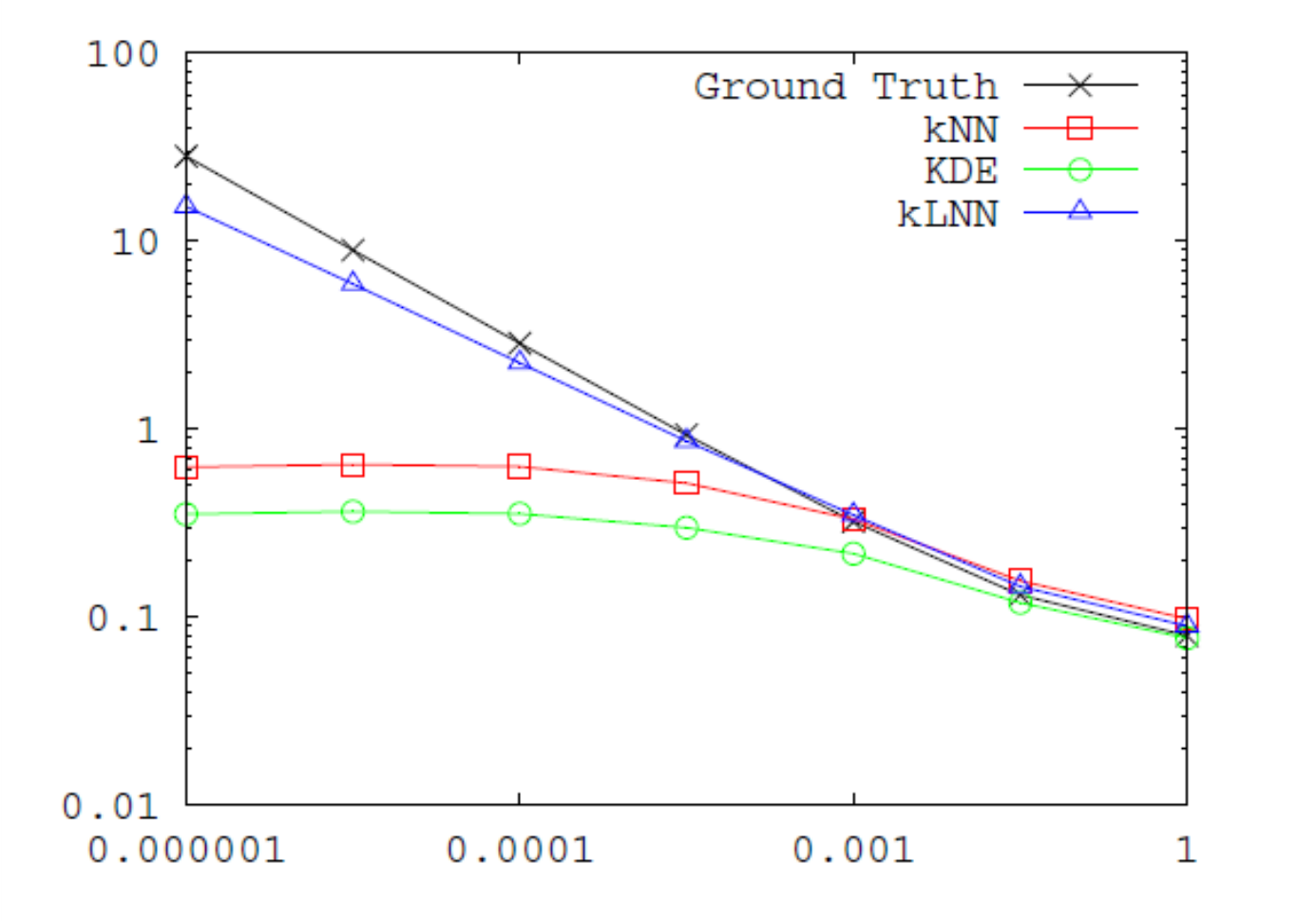}
	\put(-240,80){\small$ \E[\hJ_2(X)] $}
	\put(-170,-5){ $(1-r)$ where $r$ is correlation}
	\hspace{0.9 cm}
	\includegraphics[width=.45\textwidth]{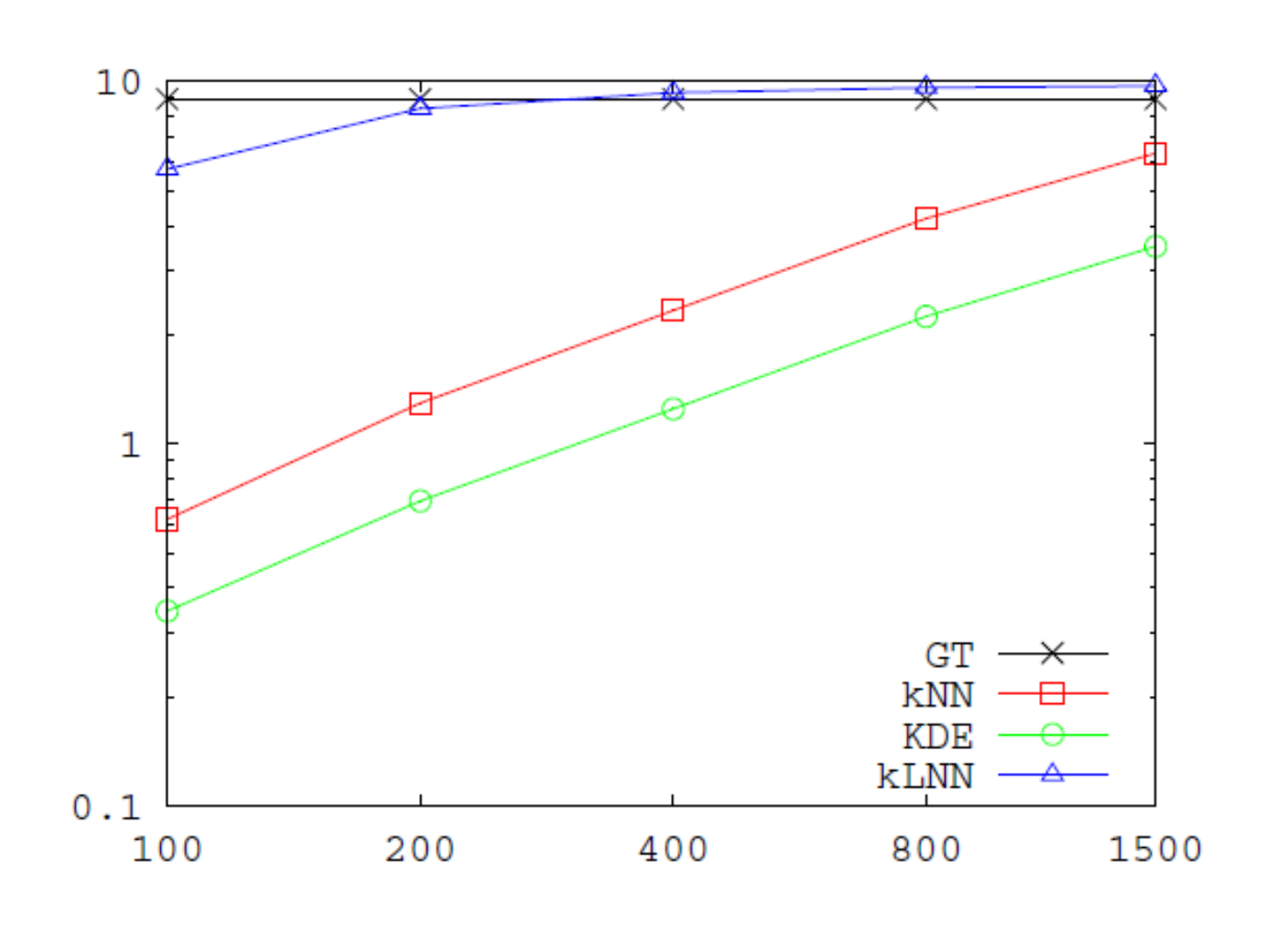}
	\put(-245,80){\small$ \E[\hJ_2(X)] $}
	\put(-150,-0){ number of samples $n$}
	\end{center}
	\caption{Proposed estimator outperform other estimators for $J_2(X)$ for mixture of highly correlated Gaussian.}
	\label{fig:mixed}
\end{figure}

In this section, we show the advantage of the $k$-LNN estimators via several synthetic experiments, by comparing it to KDE based estimators and $k$-NN based estimators~\cite{Leo08}.
%From the results, we can conclude the following advantages of the proposed $k$-LNN estimator:
%\begin{itemize}
%    \item
In the left panels in figures \ref{fig:d=1_alpha=2}--\ref{fig:mixed}, we experiment on  distributions that have very sharp boundaries ($r$ is close to 1).
Both KDE and $k$-NN based estimator fail to estimate $J_{\alpha}(X)$ accurately, whereas $k$-LNN estimator is able to reduce the boundary bias and give a better estimate. This advantage holds for different $\alpha$, for both low-dimensional  and high-dimensional spaces and for both Gaussian and non-Gaussian distributions; we conclude that the improvement is universal.

%    \item
The right panels in figures \ref{fig:d=1_alpha=2}--\ref{fig:mixed} show 
that both KDE and $k$-NN based estimators  asymptotically converge to the ground truth. But the convergence rate is much slower than $k$-LNN estimator which can provide a reasonably good estimate from small dataset. Further, note that the advantage in convergence rate holds for different $\alpha$, dimension and underlying distribution.
%\end{itemize}

%\subsection*{Experiment I: Highly Correlated Joint Gaussian}
%\label{subsec:simul1}
\bigskip\noindent{\bf Experiment I: Highly Correlated Joint Gaussian.}
Consider $X \sim \mathcal{N}\left((0,0), \begin{pmatrix} 1 & r \\ r & 1\end{pmatrix}\right)$, where the correlation $r$ is closed to 1. We estimate $J_2(X) = \int f^2(x) dx$, where the ground truth is $1/(4\pi\sqrt{1-r^2})$. In this case, the density function $f$ varies dramatically in the neighborhood of almost every point $x$. Hence, the KDE based estimator and $k$-NN based estimator suffer from boundary bias, whereas our estimator performs better. The result is shown in Figure~\ref{fig:d=1_alpha=2}. For all the experiments in this section, in the left figure, we draw 100 i.i.d.\ samples from distributions of different $r$ and plot the performance of estimators against $r$ and in the right figure, we fixed $r = 0.99999$ and show the performance against number of samples. All results are averaged over 100 independent trails.
\\

\bigskip\noindent{\bf Experiment II: Cubic Function.}
Now we consider estimation of the integral of cubic function of density $J_3(X) = \int f^3(x) dx$, where the underlying distribution is the same as in experiment I. The ground truth is $J_3(X) = 1/(12\pi^2(1-r^2))$. The result is shown in Figure~\ref{fig:d=1_alpha=3}.
\\

\bigskip\noindent{\bf Experiment III: High Dimension.}
We consider a 6-dimensional joint Gaussian random variable with ${\rm Cov}(X_1,X_2) = {\rm Cov}(X_3,X_4) = {\rm Cov}(X_5,X_6) = r$ and ${\rm Cov}(X_i, X_j) = 0$ for all other pairs of $(i,j)$. Also the integral of quadratic function $J_2(X)$ is considered. This is a generalization of experiment I for higher dimension. The result is shown in Figure~\ref{fig:d=3_alpha=2}.
\\

\bigskip\noindent{\bf Experiment IV: Mixture of Gaussian.}
We consider a non-Gaussian distribution. Let $X$ be a mixture of $\mathcal{N}\left((0,0), \begin{pmatrix} 1 & r \\ r & 1\end{pmatrix}\right)$ and $\mathcal{N}\left((0,0), \begin{pmatrix} 1 & -r \\ -r & 1\end{pmatrix}\right)$, with probability $1/2$ each. Also we consider $J_2(X)$. The result is shown in Figure~\ref{fig:mixed}.
\\

% ---------------------------------------------------------------------------------------------------------------------------------
\section{Discussion}
\label{sec:disc}

The problem of estimating integral  functional of densities has been studied for decades. 
 The minimax lower bound for the convergence rate has been established in~\cite{BM95}, and 
 several approaches have been proposed to achieve the minimax optimal rate, including Haar wavelet method~\cite{Kerk96}, Lepski's method~\cite{Mukh15} and ensemble methods~\cite{Moon16,BSY16}. 
 %However, the lower bound and all the achievability results are based on the assumption that the density $f$ is smooth (precisely, $f$ is assumed to be H\'{o}lder continuous). %For densities with sharp boundaries, little theoretical result of the minimax optimal rate is known.
%Characterizing the minimax optimal rate for non-smooth densities is an interesting open problem. 
It is unlikely that the proposed estimator will achieve the minimax rate. 
However, given its superior performance in the finite sample regime, especially for densities with sharp boundaries, 
understanding  the convergence rate of the bias for the proposed $k$-NN bandwidth estimators is an interesting open problem. 
% can shine a new light on 
%achievable rates with non-smooth densities. 
%The key challenge is to propose a class of function which can model the non-smoothness of density. Under this function class, one could be able to construct a lower bound related to the non-smoothness parameter, as well as explain the empirical performance of $k$-LNN estimator. Possibly one could also prove that the $k$-LNN estimator is minimax optimal, or propose a novel estimator which is minimax optimal. 

\section{Proof of Theorem \ref{thm:unbiased_KDE}}
\label{sec:proof_KDE}

\subsection{Proof of Asymptotic Unbiasedness}
We rewrite the estimate as
\begin{eqnarray*}
	\hJ^{{\rm (KDE)}}_{\alpha}=\frac{1}{nB_{k,d,\alpha,K}} \sum_{i=1}^n \Big\{ \underbrace{\left(\, h\big(\,(c_d nf(X_i))^{1/d} Z_{k,i}, S_{0,i}\,\big) f(X_i) \,\right)^{\alpha-1}}_{\equiv J_i}  \Big\} \;,
\end{eqnarray*}
where $S_{0,i} = \sum_{j \in \cS_{i,m}} K((X_j-X_i)/\|Z_{k,i}\|)$ and $h(t_1, t_2) = c_d t_2/\|t_1\|^d$. Since the random variables $J_1, J_2, \dots, J_n$ are identically distributed, the expected value of $\hJ_{\alpha}^{{\rm (KDE)}}$ is equal to
\begin{eqnarray}
    \E[\hJ_{\alpha}^{{\rm (KDE)}}] &=& \frac{1}{B_{k,d,\alpha,K}}\E[J_1]  \;\;= \;\; \frac{1}{B_{k,d,\alpha,K}}\E_{X_1}\big[\E[J_1|X_1 = x]\big]\;\;\,
\end{eqnarray}

If we take the limit that $n$ goes to infinity, typical approach of dominated convergence theorem cannot be used to handle the above limit. In order to exchange the limit with the (conditional) expectation over $X_1$, we assume the following Ansatz \ref{ansatz} to be true.
\begin{ansatz}
    \label{ansatz}
	The function $h(\cdot,\cdot)$ is bounded.
\end{ansatz}

As noted in \cite{Pal10} this ansatz is commonly used implicitly in the literature on consistency of $k$-NN estimators, without explicitly stating as such, in existing analyses of consistency of entropy estimators including \cite{KL87,GLMN05,Leo08,WKV09}.
This assumption can be avoided for results of the convergence rate of the estimator with respect to the sample size with more assumptions as in \cite{Pal10,GOV16De,SP16,BSY16}. In practice, we can truncated $h$ by some very large constants to fulfill the ansatz.

Under this ansatz, by dominant convergence theorem, we can exchange the limit with the conditional expectation and obtain
\begin{eqnarray}
    \label{eq:entropy1}
\lim_{n \to \infty} \E[\hJ_{\alpha}^{{\rm (KDE)}}] = \frac{1}{B_{k,d,\alpha,K}} \E_{X_1} \left[\, \lim_{n \to \infty} \E \left[\, J_1 | X_1 = x \,\right] \,\right].
\end{eqnarray}

Now we will show that the expectation inside converges to $(f(x))^{\alpha-1}$ multiplied by some constant that is independent of the underlying distribution. Precisely, for almost every $x$ and given $X_1 = x$, we have
\begin{eqnarray}
\E[J_1 | X_1 = x] &=& \E \left[\, \left(\, h ( (c_d nf(x))^{1/d} Z_{k,1}, S_{0,1}) f(x)\,\right)^{\alpha-1} \,\right] \,\notag\\
& \longrightarrow & B_{k,d,\alpha,K} (f(x))^{\alpha-1}\;,
\label{eq:entropy2}
%\E\left[\, \bar{h} \left(\,\xi_k \sum_{l=1}^k E_\ell, \tilde{S}^{(\infty)}_{0,i}, \tilde{S}^{(\infty)}_{1,i}, \tilde{S}^{(\infty)}_{2,i} \,\right) \,\right] \;,
\end{eqnarray}
as $n\to \infty$. Here $B_{k,d,\alpha.K}$ is a constant only depends on $k$ $d$,$\alpha$ and $K$, defined in \eqref{eq:defB}. Therefore,
\begin{eqnarray}
    \E_{X_1} \left[\, \lim_{n \to \infty} \E[J_1|X_1 = x]\right] & =&     \E_{X_1} [B_{k,d,\alpha,K} (f(X_1))^{\alpha-1}] \,\notag\\
    &=& B_{k,d,\alpha,K} J_{\alpha}(X) \;.\label{eq:entropy4}
\end{eqnarray}

Together with \eqref{eq:entropy1}, this finishes the proof of the desired claim.
\\

We are now left to prove the convergence of  \eqref{eq:entropy2}. We first give a formal definition of the multiplicative factor $B_{k,d,\alpha,K}$ by replacing the sample defined quantities $S_{0,1}$ by similar quantities defined by order statistics,
and use Lemma~\ref{lem:order_stat} to prove the convergence.
Recall that our order statistics is defined by two sequences of $m$ i.i.d. random variables:
i.i.d. standard exponential random variables $E_1, \dots, E_m$ and
i.i.d. Haar random variables  $\xi_1, \dots, \xi_m$ uniformly distributed over $d$-dimensional unit sphere. Now we define
\begin{eqnarray}
	\label{eq:defB}
	B_{k,d,\alpha,K} &\equiv &
	\E\left[\, \left( h \left(\, \xi_k (\sum_{\ell=1}^k E_\ell)^{1/d}, \tilde{S}_0^{(\infty)} \,\right) \,\right)^{\alpha-1}\,\right] \;,
\end{eqnarray}
here $\tS_0^{(\infty)}$ is defined by the limit of a convergent random sequence
\begin{eqnarray}
\tilde{S}_0^{(m)} &\equiv&  \sum_{j=1}^{m} K\left(\, \frac{\xi_j (\sum_{\ell=1}^j E_\ell)^{1/d} }{(\sum_{\ell=1}^k E_\ell)^{1/d} } \,\right) \;,
\end{eqnarray}

We will show that the limit exists in Lemma~\ref{lem:tail}. We introduce simpler notations for the joint random variables:
$\tS^{(m)}=(\xi_k (\sum_{\ell=1}^k E_{\ell})^{1/d}, \tilde{S}^{(m)}_{0})$ and
 $\tS^{(\infty)}=(\xi_k (\sum_{\ell=1}^k E_{\ell})^{1/d}, \tilde{S}^{(\infty)}_{0})$.
Considering the quantities  $S^{(n)} = ((c_d nf(x))^{1/d} Z_{k,1}, S_{0,1})$
defined from samples, we show that this converges
to $\tS^{(\infty)}$. Precisely, by applying triangular inequality,
\begin{eqnarray}
	d_{\rm TV} (S^{(n)},\tS^{(\infty)}) &\leq&
	d_{\rm TV} (S^{(n)},\tS^{(m)}) + d_{\rm TV} (\tS^{(m)},\tS^{(\infty)})  \;, \label{eq:entropy3}
\end{eqnarray}
and we show that both terms converge to zero for any $m = \Theta(\log n)$.
Given that $\bh$ is continuous and bounded from the ansatz, we obtain
\begin{eqnarray}
	\lim_{n\to\infty} \E[J_1|X_1 = x] &=& \E\, \left[\, \lim_{n\to\infty} \left(\, h(S^{(n)}) f(x) \,\right)^{\alpha-1}|X_1 =  x \,\right] \,\notag\\
	&=& (f(x))^{\alpha-1} \E \,\left[ \,(h(\tS^{(\infty)}))^{\alpha-1} \,\right]\;,
\end{eqnarray}
for almost every $x$, proving \eqref{eq:entropy4}.

The convergence of the first term follows from Lemma \ref{lem:order_stat}.
Precisely, consider the function $g_{m}: \mathbb{R}^{d \times m} \to \mathbb{R}^d \times \mathbb{R} $ defined as:
\begin{eqnarray}
g_{m}(t_1, t_2, \dots, t_{m}) = \left(\, t_k, \sum_{j=1}^{m} K \left(\, \frac{t_j}{\|t_k\|} \,\right) \,\right) \;,
\end{eqnarray}
such that $S^{(n)} = g_{m} \left(\, (c_dnf(x))^{1/d} \left(\, Z_{1,i}, Z_{2,i}, \dots, Z_{m,i} \,\right) \,\right)$, which follows from the definition of
$S^{(n)}=((c_dnf(x))^{1/d}Z_{k,i},S_{0,i})$. Similarly, $\tS^{(m)} = g_{m} \left(\, \xi_1 E_1^{1/d}, \xi_2(E_1 + E_2)^{1/d}, \dots \xi_{m}(\sum_{\ell=1}^{m} E_\ell)^{1/d} \,\right)$. Since $g_{m}$ is continuous, so for any set $A \in \mathbb{R}^d \times \mathbb{R}$, there exists a set $\tA \in \mathbb{R}^{d \times m}$ such that $g_{m}(\tA) = A$. So for any $x$ such that there exists $\varepsilon > 0$ such that $f(a) > 0$, $\|\nabla f(a)\| = O(1)$ and $\|H_f(a)\| = O(1)$ for any $\|a - x\| < \varepsilon$, we have:
\begin{align}
& d_{\rm TV} ( S^{(n)},\tS^{(m)})  \,\notag\\
&= \sup_{A } \left|\, \Pr\left\{g_{m} \left(\, (c_dnf(x))^{1/d} Z_{1,i},  \dots, (c_dnf(x))^{1/d}Z_{m,i} \,\right) \in A \right\} - \Pr\{g_{m} (\, \xi_1 E_1^{1/d}, \dots \xi_{m}(\sum_{l=1}^{m} E_{\ell})^{1/d} \,) \in A \} \,\right| \,\notag\\
&\leq \sup_{\tA \in \mathbb{R}^{d \times m}} \left|\, \Pr\left\{\left(\, (c_dnf(x))^{1/d} Z_{1,i},  \dots, (c_dnf(x))^{1/d} Z_{m,i} \,\right) \in \tA \right\} - \Pr\{(\, \xi_1 E_1^{1/d},  \dots \xi_{m} (\sum_{\ell=1}^{m} E_\ell)^{1/d} \,) \in \tA \} \,\right| \,\notag\\
&= d_{\rm TV}\left( \left(\, (c_dnf(x))^{1/d} Z_{1,i},  \dots, (c_dnf(x))^{1/d} Z_{m,i} \,\right) \,,\, \left(\, \xi_1 E_1^{1/d}, \dots \xi_{m} (\sum_{\ell=1}^{m} E_\ell)^{1/d} \,\right)\,\right)  \,\notag\\
	&\stackrel{n \rightarrow \infty}{\longrightarrow} 0 \label{eq:converge_1}\;,
\end{align}
where the last inequality follows from Lemma \ref{lem:order_stat}. By the assumption that $f$ has open support and $\|\nabla f\|$ and $\|H_f\|$ is bounded almost everywhere, this convergence holds for almost every $x$.
\\

For the second term in \eqref{eq:entropy3}, let $\tilde{T}^{(m)}_0 = \tS^{(\infty)}_0 - \tS^{(m)}_0$ and we claim that $\tilde{S}^{(m)}$ converges to $\tilde{S}^{(\infty)}$ in distribution by the following lemma.
\begin{lemma}
    \label{lem:tail}
    Assume $m_n \to \infty$ as $n \to \infty$, and the kernel functions $K: \mathbb{R}^d \to \mathbb{R}^{d'}$ satisfied $\|K(u)\|  \leq C \|u\|^{-2d}$ for some constant $C > 0$. Then we have
    \begin{eqnarray}
    \lim_{n \to \infty} \E \,\Big\|\, \sum_{j=m_n+1}^{\infty} K\left(\, \frac{\xi_j (\sum_{\ell=1}^j E_{\ell})^{1/d}}{(\sum_{\ell=1}^k E_{\ell})^{1/d}}\,\right)\,\Big\| = 0 \;.
    \end{eqnarray}
\end{lemma}
This implies that $\tilde{T}_0^{(m)}$ converges to $0$ in $L_1$. Therefore $\tS^{(m)}=(\xi_k (\sum_{\ell=1}^k E_{\ell})^{1/d}, \tilde{S}^{(m)}_{0})$ converges to $\tS^{(\infty)}=(\xi_k (\sum_{\ell=1}^k E_{\ell})^{1/d}, \tilde{S}^{(\infty)}_{0})$ in $L_1$, hence, in distribution. Therefore,
\begin{eqnarray}
d_{\rm TV}(\tS^{(m)},\tS^{(\infty)}) \stackrel{n \rightarrow \infty}{\longrightarrow} 0 \label{eq:converge_2}\;,
\end{eqnarray}

Combine~\eqref{eq:converge_1} and~\eqref{eq:converge_2} in~\eqref{eq:entropy3}, this implies the desired claim.

\subsection{Proof of the Variance}
We will follow the technique from \cite[Section 7.3]{BD16}.
For the usage of Efron-Stein inequality, we need a second set of i.i.d. samples $\{X'_1, X'_2, \dots, X'_n\}$. For simplicity, denote $\hJ = \hJ_{\alpha}^{{\rm (KDE)}}(X)$ be the estimate of $J(X)$ base on original sample $\{X_1, \dots, X_n\}$ and $\hJ^{(i)}$ be the estimate based on $\{X_1, \dots, X_{i-1}, X'_i, X_{i+1}, \dots X_n\}$, where only $X_i$ is replaced by $X'_i$. Then Efron-Stein theorem states that
\begin{eqnarray}
\Var \left[ \hJ \right] \leq 2 \sum_{j=1}^n  \mathbb{E} \left[\, \left( \hJ - \hJ^{(j)} \right)^2 \,\right] \;.\label{eq:efron_stein}
\end{eqnarray}

Recall that
\begin{eqnarray*}
	\hJ^{(n)}_{\alpha}=\frac{1}{nB_{k,d,\alpha,K}} \sum_{i=1}^n \Big\{ \underbrace{\left(\, h\big(\,(c_d nf(X_i))^{1/d} Z_{k,i}, S_{0,i}\,\big) f(X_i) \,\right)^{\alpha-1}}_{\equiv J_i}  \Big\} \;,
\end{eqnarray*}

Similarly, we can write $\hJ^{(j)} = (1/nB_{k,d,\alpha,K}) \sum_{i=1}^n J_i^{(j)}$ for any $j \in \{1, \dots, n\}$. Therefore, the difference of $\hJ$ and $\hJ^{(j)}$ is
\begin{eqnarray}
 \hJ - \hJ^{(j)} = \frac{1}{nB_{k,d,\alpha,K}} \sum_{i=1}^n \left(\, J_i - J_i^{(j)} \,\right) \;.
\end{eqnarray}

Notice that $J_i$ only depends on $X_i$ and its $m$ nearest neighbors, so $J_i - J_i^{(j)} = 0$ if none of $X_j$ and $X'_j$ are in $m$ nearest neighbor of $X_i$. If we denote $Z_{i,j} = \mathbb{I} \{X_j \textrm{ is in } m \textrm{ nearest neighbor of } X_i\}$, then $J_i = J_i^{(j)}$ if $Z_{i,j}+Z_{i,j'} = 0$. According to \cite[Lemma 20.6]{BD16}, since $X$ has a density, with probability one, $\sum_{i=1}^n Z_{i,j} \leq m \gamma_d$, where $\gamma_d$ is the minimal number of cones of angle $\pi/6$ that can cover $\mathbb{R}^d$, which only depends on $d$. Similarly, $\sum_{i=1}^n Z_{i,j'} \leq m \gamma_d$. If we denote $S_j = \{i: Z_{i,j} + Z_{i,j'} > 0\}$, the cardinality of $S$ satisfy $|S_j| \leq 2 m \gamma_d$. Therefore, we have $\hJ - \hJ^{(j)} = \sum_{i \in S} \left(\, J_i - J_i^{(j)} \,\right)/(nB_{k,d,\alpha,K})$. By Cauchy-Schwarz inequality, we have
\begin{eqnarray}
\mathbb{E} \left[\, \left( \hJ - \hJ^{(j)} \right)^2 \,\right] &=& \mathbb{E} \left[\, \frac{1}{n^2 B_{k,d,\alpha,K}^2} \left(\, \sum_{i \in S_j} \left(\, J_i - J_i^{(j)} \,\right) \,\right)^2\,\right] \,\notag\\
&\leq& \mathbb{E} \left[\, \frac{|S_j|}{n^2 B_{k,d,\alpha,K}^2} \sum_{i \in S_j} \left(\, J_i - J_i^{(j)} \,\right)^2\,\right] \,\notag\\
&=& \frac{|S_j|}{n^2 B_{k,d,\alpha,K}^2} \sum_{i \in S_j} \mathbb{E} \left[\, \left(\, J_i - J_i^{(j)}\,\right)^2\,\right] \,\notag\\
&\leq& \frac{2|S_j|}{n^2 B_{k,d,\alpha,K}^2} \sum_{i \in S_j} \left(\, \mathbb{E} \left[\, J_i^2\,\right] + \mathbb{E} \left[\, (J_i^{(j)})^2\,\right] \,\right) \;.\label{eq:h-h_j}
\end{eqnarray}
for every $j \in [n]$. Notice that $J_i$'s and $J_i^{(j)}$'s are identically distributed, so we are left to compute $\mathbb{E} \left[\, J_1^2 \,\right]$. Conditioning on $X_1 = x$, similarly to~\eqref{eq:entropy2}, we have
\begin{eqnarray}
\E[J_1^2 | X_1 = x] &=& \E \left[\, \left|\, h ( (c_d nf(x))^{1/d} Z_{k,i}, S_{0,1}) f(x)\,\right|^{2\alpha-2} \,\right] \,\notag\\
& \longrightarrow & B_{k,d,2\alpha-1,K} |f(x)|^{2\alpha-2}\;,
\end{eqnarray}
as $n \to \infty$. Therefore, by taking expectation over $X_1$, we obtain:
\begin{eqnarray}
\E [J_1^2] &=& \E_{X_1} \left[\, \lim_{n \to \infty} \E \left[\, J_1^2 | X_1 \,\right] \,\right] = B_{k,d,2\alpha-1,K} \E_{X_1} \left[\, |f(X_1)|^{2\alpha-2} \,\right] < +\infty\;,
\end{eqnarray}
where the last inequality comes from the assumption that $\E \left[\, |f(X)|^{2\alpha-2} \,\right] < +\infty$. Combining with~\eqref{eq:efron_stein} and~\eqref{eq:h-h_j}, we have
\begin{eqnarray}
\Var \left[ \hJ \right] &\leq& 2 \sum_{j=1}^n  \mathbb{E} \left[\, \left( \hJ - \hJ^{(j)} \right)^2 \,\right] \,\notag\\
&\leq& \frac{4}{n^2 B^2_{k,d,\alpha,K}} \sum_{j=1}^n \left(\, |S_j| \sum_{i \in S_j} \left(\, \mathbb{E} \left[\, J_i^2\,\right] + \mathbb{E} \left[\, (J_i^{(j)})^2\,\right] \,\right) \,\right) \,\notag\\
&\leq& \frac{4}{n^2B^2_{k,d,\alpha,K}} \sum_{j=1}^n \left(\, 2 |S_j|^2 B_{k,d,2\alpha-1,K} C\,\right) \leq \frac{32m^2 \gamma_d^2 B_{k,d,2\alpha-1,K} C}{n B^2_{k,d,\alpha,K}} \;,
\end{eqnarray}
where $C$ is the upper bound for $\E \left[\, |f(X)|^{2\alpha-2} \,\right]$. Take $m = O(\log n)$ then the proof is complete.

\subsection{Proof of Corollary~\ref{cor:renyi_unbiased_KDE}}

For any positive real number $\epsilon > 0$, we have
\begin{eqnarray}
&&\Pr \left(\, |\hH_{\alpha}^{{\rm (KDE)}}(X) - H_{\alpha}(X)| > \epsilon\,\right) \,\notag\\
&=& \Pr \left(\, |\frac{1}{1-\alpha} \left(\, \log \hJ_{\alpha}^{{\rm (KDE)}}(X) - \log J_{\alpha}(X) \,\right)| > \epsilon\,\right) \,\notag\\
&=& \Pr \left(\, | \log \hJ_{\alpha}^{{\rm (KDE)}}(X) - \log J_{\alpha}(X) | > \epsilon|1-\alpha| \,\right) \,\notag\\
&=& \Pr \left(\, \hJ_{\alpha}^{{\rm (KDE)}}(X) > J_{\alpha}(X)  e^{\epsilon|1-\alpha|} \,\right) + \Pr \left(\, \hJ_{\alpha}^{{\rm (KDE)}}(X) < J_{\alpha}(X)  e^{-\epsilon|1-\alpha|} \,\right) \,\notag\\
&=& \Pr \left(\, \hJ_{\alpha}^{{\rm (KDE)}}(X) - J_{\alpha}(X) > J_{\alpha}(X)  (e^{\epsilon|1-\alpha|}-1) \,\right) + \Pr \left(\, \hJ_{\alpha}^{{\rm (KDE)}}(X) - J_{\alpha}(X) < J_{\alpha}(X)  (e^{-\epsilon|1-\alpha|}-1) \,\right) \,\notag\\
&\leq& \frac{\E\left[\,\left(\,\hJ_{\alpha}^{{\rm (KDE)}}(X) - J_{\alpha}(X)\,\right)^2\,\right]}{J_{\alpha}^2(X)(e^{\epsilon|1-\alpha|}-1)^2} + \frac{\E\left[\,\left(\,\hJ_{\alpha}^{{\rm (KDE)}}(X) - J_{\alpha}(X)\,\right)^2\,\right]}{J_{\alpha}^2(X)(1-e^{-\epsilon|1-\alpha|})^2}
\end{eqnarray}
where the last inequality is Chebyshev inequality. Since $\epsilon$, $\alpha$ and $J_{\alpha}(X)$ are all fixed quantities, and $ \E\left[\,\left(\,\hJ_{\alpha}^{{\rm (KDE)}}(X) - J_{\alpha}(X)\,\right)^2\,\right] \to 0$ as $n$ tends to infinity, as shown in Theorem~\ref{thm:unbiased_KDE}. Therefore, the probability $\Pr \left(\, |\hH_{\alpha}^{{\rm (KDE)}}(X) - H_{\alpha}(X)| > \epsilon\,\right)$ vanishes as $n \to \infty$, i.e., $\hH_{\alpha}^{{\rm (KDE)}}(X)$ converges to $H_{\alpha}(X)$ in probability.

\subsection{Proof of Lemma~\ref{lem:tail}}

Firstly, since $\|K(u)\| \leq C \|u\|^{-2d}$ for all $u$, we can upper bound the expectationby:
\begin{eqnarray}
&& \E \, \Big\|\, \sum_{j=m_n+1}^{\infty} K\left(\, \frac{\xi_j(\sum_{l=1}^j E_l)^{1/d}}{(\sum_{l=1}^k E_l)^{1/d}} \,\right)\Big\|\, \,\notag\\
&\leq& \sum_{j=m_n+1}^{\infty} \E\, \Big\|\, K\left(\, \frac{\xi_j(\sum_{l=1}^j E_l)^{1/d}}{(\sum_{l=1}^k E_l)^{1/d}} \,\right)\,\Big\|\,\notag\\
&\leq& C \sum_{j=m_n+1}^{\infty} \E \, \Big\| \frac{\xi_j(\sum_{l=1}^j E_l)^{1/d}}{(\sum_{l=1}^k E_l)^{1/d}} \Big\|^{-2d} \,\notag\\
& = & C \sum_{j=m_n+1}^{\infty} \E\, \left[\, \frac{(\sum_{l=1}^k E_l)^{2}}{(\sum_{l=1}^j E_l)^{2}} \,\right]\label{eq:eq1}
\end{eqnarray}
where the last equality comes from the fact that $\|\xi_j\| = 1$ for all $j$. Now for any fixed $j \geq k$, let $R_k = \sum_{l=1}^k E_l$ and $R_{j-k} = \sum_{l=k+1}^j E_l$. Notice that $R_k$ is the summation of $k$ i.i.d. standard exponential random variables, so $R_k \sim \textit{Erlang} (k,1)$. Similarly, $R_{j-k} \sim \textit{Erlang}(j-k,1)$. Also $R_k$ and $R_{j-k}$ are independent. Recall that the pdf of $\textit{Erlang}(k, \lambda)$ is given by $f_{k,\lambda}(x) = \lambda^k x^{k-1} e^{-\lambda}/(k-1)!$ for $x \geq 0$. So we have:
\begin{eqnarray}
&&\E\, \left[\, \frac{(\sum_{l=1}^k E_l)^{2}}{(\sum_{l=1}^j E_l)^{2}} \,\right] = \E \left[\, \frac{R_k^2}{(R_k+R_{j-k})^2} \,\right] \,\notag\\
&=& \int_{x,y \geq 0} \frac{x^2}{(x+y)^2} \frac{x^{k-1}e^{-x}}{(k-1)!} \frac{y^{j-k-1}e^{-y}}{(j-k-1)!} dx dy \,\notag\\
&\leq& \int_{x,y \geq 0} \frac{x^2}{(x+y)^2} \frac{x^{k-1}e^{-x}}{(k-1)!} \frac{y^{j-k-3}(x+y)^2 e^{-y}}{(j-k-1)!} dx dy \,\notag\\
&=& \int_{x,y \geq 0} \frac{x^{k+1}e^{-x}}{(k-1)!} \frac{y^{j-k-3}e^{-y}}{(j-k-1)!} dx dy \,\notag\\
&=& \frac{(k+1)!}{(k-1)!} \frac{(j-k-3)!}{(j-k-1)!} = \frac{k(k+1)}{(j-k-1)(j-k-2)}\label{eq:Ej}\;.
\end{eqnarray}\
Therefore, for sufficiently large $n$ such that $m_n \geq 2k+4$, i.e., $m_n - k - 2 \geq m_n/2$, we have
\begin{eqnarray}
&&\E \, \Big\|\, \sum_{j=m_n+1}^{\infty} K\left(\, \frac{\xi_j(\sum_{l=1}^j E_l)^{1/d}}{(\sum_{l=1}^k E_l)^{1/d}} \,\right)\Big\| \leq C \sum_{j=m_n+1}^{\infty} \E\, \left[\, \frac{(\sum_{l=1}^k E_l)^{2}}{(\sum_{l=1}^j E_l)^{2}} \,\right] \,\notag\\
&\leq& C \sum_{j=m_n+1}^{\infty} \frac{k(k+1)}{(j-k-1)(j-k-2)} \,\notag\\
&=& Ck(k+1) \sum_{j=m_n+1}^{\infty} (\frac{1}{j-k-2} - \frac{1}{j-k-1}) = \frac{Ck(k+1)}{m_n-k-1} \;.
\end{eqnarray}

Notice that $m_n \to \infty$ as $n \to \infty$, therefore,
\begin{eqnarray}
\lim_{n \to \infty} \E \,\Big\|\, \sum_{j=m_n+1}^{\infty} K\left(\, \frac{\xi_j(\sum_{l=1}^j E_l)^{1/d}}{(\sum_{l=1}^k E_l)^{1/d}} \,\right)\Big\| = 0 \;.
\end{eqnarray}

\section{Proof of Theorem \ref{thm:unbiased_kLNN}}
\label{sec:proof_kLNN}

The proof is quite similar to the proof of Theorem~\ref{thm:unbiased_KDE}, so we skip the detail and focus on the main steps below. First, we rewrite the estimator as
\begin{eqnarray*}
	\hJ^{(k-{\rm LNN})}_{\alpha}=\frac{1}{nB_{k,d,\alpha}} \sum_{i=1}^n \Big\{ \underbrace{\left(\, h\big(\,(c_d nf(X_i))^{1/d} Z_{k,i}, S_{0,i}, S_{1,i}, S_{2,i})\,\big) f(X_i) \,\right)^{\alpha-1}}_{\equiv J_i}  \Big\} \;,
\end{eqnarray*}
here the quantities $S_{0,i}$, $S_{1,i}$, $S_{2,i}$ and $\mu_i$, $\Sigma_i$ are given as follows,
\begin{eqnarray}
\label{eq:defS_i}
S_{0,i} &\equiv& \sum_{j \in \cS_{i,m}} e^{-\frac{\|X_j - X_i\|^2}{2\rho_{k,i}^2}} \;,\\
S_{1,i} &\equiv& \sum_{j \in \cS_{i,m}} \frac{X_j - X_i}{\rho_{k,i}} \, e^{-\frac{\|X_j - X_i\|^2}{2\rho_{k,i}^2}} \;,\\
S_{2,i} &\equiv& \sum_{j \in \cS_{i,m}} \frac{(X_j - X_i)(X_j - X_i)^T}{\rho_{k,i}^2} \, e^{-\frac{-|X_j - X_i\|^2}{2\rho_{k,i}^2}} \;,\\
\mu_i &\equiv& \frac{S_{1,i}}{S_{0,i}} \;,\\
\Sigma_i &\equiv& \frac{S_{2,i}}{S_{0,i}} - \frac{S_{1,i} S_{1,i}^T}{S_{0,i}^2} \;.
\end{eqnarray}
and $h : \mathbb{R}^d \times \mathbb{R} \times \mathbb{R}^d \times \mathbb{R}^{d \times d} \to \mathbb{R} $ is defined as
\begin{align}
&h(t_1, t_2, t_3, t_4) = \frac{C_d t_2}{\|t_1\|^d (2\pi)^{d/2} \det\left(\, \frac{t_4}{t_2} - \frac{t_3 t_3^T}{t^2_2}\,\right)^{1/2}} \exp\{-\frac{1}{2} t_3^T (t_2t_4 - t_3 t_3^T)^{-1} t_3\}\;.
\end{align}

Since $J_1, J_2, \dots, J_n$ are identically distributed, we have $\E \left[\, \hJ^{(k-{\rm LNN})}_{\alpha}\,\right] = \E_{X_1} [ \E [J_1 | X_1 = x] ] /B_{k,d,\alpha}$. By assuming the ansatz that $h(\cdot,\cdot,\cdot,\cdot)$ is bounded, we are able to exchange the limit and conditional expectation, therefore, we are left to show that

\begin{eqnarray}
\E[J_1 | X_1 = x] &=& \E \left[\, \left(\, h ( (c_d nf(x))^{1/d} Z_{k,i}, S_{0,1}, S_{1,i}, S_{2,i}) f(x)\,\right)^{\alpha-1} \,\right] \,\notag\\
& \longrightarrow & B_{k,d,\alpha} (f(x))^{\alpha-1}\;,
\label{eq:entropy2_L}
\end{eqnarray}

To prove this, we show that the empirical quantities $((c_d n f(x))^{1/d} Z_{k,1}, S_{0,1}, S_{1,1}, S_{2,1})$ jointly converges to $(\xi_k (\sum_{\ell=1}^k E_{\ell})^{1/d}, \tS_0^{(\infty)}, \tS_1^{(\infty)}, \tS_2^{(\infty)})$ in distribution. Here $\tS_{\gamma}^{(\infty)}$ is defined by the limit of the following convergent random sequence
\begin{eqnarray}
\tilde{S}^{(m)}_{\gamma}  &\equiv&  \sum_{j=1}^{m} \frac{\xi_j^{(\gamma)} (\sum_{\ell=1}^j E_\ell)^{\gamma/d} }{(\sum_{\ell=1}^k E_\ell)^{\gamma/d} } \exp\Big\{- \frac{(\,\sum_{\ell=1}^j E_\ell\,)^{2/d}}{2(\,\sum_{\ell=1}^k E_\ell \,)^{2/d}} \Big\} \;,
\end{eqnarray}
where $\xi_j^{(0)} = 1$, $\xi_j^{(1)} = \xi_j$, $\xi_j^{(2)} = \xi_j\xi_j^T$ and $\tilde{S}^{(\infty)}_{\gamma} = \lim_{m \to \infty} \tilde{S}^{(m)}_{\gamma}$. Here Lemma~\ref{lem:order_stat} and Lemma~\ref{lem:tail} (by applying $K_0(u) = \exp\{-\|u\|^2/2\}$, $K_1(u) = u \exp\{-\|u\|^2/2\}$ and $K_2(u) = uu^T \exp\{-\|u\|^2/2\}$ for $S_{0,1}$, $S_{1,1}$ and $S_{2,1}$ respectively) are used to prove the convergence following the same approach as in the proof of Theorem~\ref{thm:unbiased_KDE}. By the assumption that $h$ is continuous and bounded, we obtain
\begin{eqnarray}
&&\lim_{n \to \infty} \E[J_1 | X_1 = x] \,\notag\\
&=& \E \left[\, \lim_{n \to \infty} \left(\, h ( (c_d nf(x))^{1/d} Z_{k,i}, S_{0,1}, S_{1,i}, S_{2,i}) f(x)\,\right)^{\alpha-1} \,\right] \,\notag\\
& =& (f(x))^{\alpha-1} \underbrace{ \E\left[\, \left( h \left(\,\xi_k \left(\, \sum_{\ell=1}^k E_\ell \,\right)^{1/d}, \tilde{S}^{(\infty)}_{0}, \tilde{S}^{(\infty)}_{1}, \tilde{S}^{(\infty)}_{2} \,\right) \,\right)^{\alpha-1} \,\right]}_{\equiv B_{K,d,\alpha}}\;.
\end{eqnarray}
which proves the asymptotic unbiasedness of $\hJ_{\alpha}^{(k-{\rm LNN})}(X)$.

For the variance, we use the Efron-Stein inequality. Let $\hJ$ be the $k$-LNN estimate of $J_{\alpha}(X)$ based on original samples and $\hJ^{(i)}$ be the estimate if $X_i$ is replaced by $X'_i$. Since the $k$-LNN estimate only uses the $m$-nearest neighbors of each sample, the set $S_j = \{i: J_i - J_i^{(j)} \neq 0\}$ has no more than $2m\gamma_d$ elements. Therefore,
\begin{eqnarray}
\Var \left[ \hJ \right] &\leq& 2 \sum_{j=1}^n  \mathbb{E} \left[\, \left( \hJ - \hJ^{(j)} \right)^2 \,\right] \,\notag\\
&\leq& \frac{4}{n^2 B^2_{k,d,\alpha}} \sum_{j=1}^n \left(\, |S_j| \sum_{i \in S_j} \left(\, \mathbb{E} \left[\, J_i^2\,\right] + \mathbb{E} \left[\, (J_i^{(j)})^2\,\right] \,\right) \,\right) \,\notag\\
&\leq& \frac{4}{n^2B^2_{k,d,\alpha}} \sum_{j=1}^n \left(\, 2 |S_j|^2 B_{k,d,2\alpha-1} C\,\right) \leq \frac{32m^2 \gamma_d^2 B_{k,d,2\alpha-1} C}{n B^2_{k,d,\alpha}} \;,
\end{eqnarray}
where $C$ is the upper bound for $\E |f(X)|^{2\alpha-2}$. Take $m = O(\log n)$ to complete the proof.

\bibliographystyle{plain}
{\small
\bibliography{renyi}}

\end{document}